\documentclass[fleqn,usenatbib]{mnras}

\usepackage{newtxtext,newtxmath}

\usepackage[T1]{fontenc}

\DeclareRobustCommand{\VAN}[3]{#2}
\let\VANthebibliography\thebibliography
\def\thebibliography{\DeclareRobustCommand{\VAN}[3]{##3}\VANthebibliography}

\usepackage{graphicx}	
\usepackage{amsmath}	

\usepackage{color}
 \usepackage{ulem}
\usepackage{lineno}

\title{Kilonovae of binary neutron star mergers leading to short-lived remnant neutron star formation} 
\author[K. Kawaguchi et al.]{
Kyohei Kawaguchi,$^{1,2,3}$\thanks{E-mail: kyohei.kawaguchi@aei.mpg.de}
Sho Fujibayashi,$^{1}$
Nanae Domoto,$^{4}$
Kenta Kiuchi,$^{1,3}$
Masaru Shibata,$^{1,3}$
Shinya Wanajo,$^{1}$
\\
$^{1}$Max Planck Institute for Gravitational Physics (Albert Einstein Institute), Am M\"{u}hlenberg 1, Potsdam-Golm, 14476, Germany\\
$^{2}$Institute for Cosmic Ray Research, The University of Tokyo, 5-1-5 Kashiwanoha, Kashiwa, Chiba 277-8582, Japan\\
$^{3}$Center for Gravitational Physics and Quantum Information, Yukawa Institute for Theoretical Physics, Kyoto University, Kyoto, 606-8502, Japan\\
$^{4}$Astronomical Institute, Tohoku University, Aoba, Sendai 980-8578, Japan
}

\date{Accepted XXX. Received YYY; in original form ZZZ}

\pubyear{20XX}

\begin{document}
\label{firstpage}
\pagerange{\pageref{firstpage}--\pageref{lastpage}}
\maketitle

\begin{abstract}
We study kilonova emission from binary neutron star (BNS) mergers for the case that a remnant massive neutron star (MNS) forms and collapses to a black hole within $20$\,ms after the onset of the merger (which we refer to as ``a short-lived case") by consistently employing numerical-relativity and nucleosynthesis results. We find that such kilonovae are fainter and last shorter than those for BNSs resulting in the formation of long-lived ($\gg 1\,{\rm s}$) MNSs, in particular in the optical band. The resulting light curves are too faint and last for a too short duration to explain the kilonova observation for the BNS associated with GW170817, indicating that the merger remnant formed in GW170817 is unlikely to have collapsed to a black hole within a short period of time ($\sim 20$\,ms) after the onset of the merger. Our present result implies that early observation is necessary to detect kilonovae associated with BNSs leading to short-lived MNS formation in particular for the optical blue band as well as that kilonovae could be hidden by the gamma-ray burst afterglow for nearly face-on observation. We provide a possible approximate scaling law for near-infrared light curves with the given reference time and magnitude when the decline power of the {\it z}-band magnitude, $d M_{\it z}/d{\rm log}_{10}t$, reaches $2.5$. This scaling law suggests that the {\it HK}-band follow-up observation should be at least $1$ mag deeper than that for the {\it z}-band reference magnitude and earlier than 4 times the reference time.
\end{abstract}




\section{Introduction}\label{sec:intro}

Binary neutron star (BNS) mergers are among the most efficient gravitational-wave emitters in the universe and the most important sources of multi-messenger high-energy astrophysical phenomena, such as gamma-ray bursts~\citep[GRB, ][]{1991AcA....41..257P,Nakar:2007yr,Berger:2013jza,LIGOScientific:2017zic}, kilonovae~\citep{Li:1998bw,Kulkarni:2005jw,Metzger:2010sy,Kasen:2013xka,Tanaka:2013ana}, and synchrotron flares~\citep{Nakar2011Natur,Hotokezaka:2015eja,Hotokezaka:2018gmo,Margalit2020MNRAS}. Furthermore, BNS mergers are considered to be important production sites of elements heavier than iron in the universe~\citep{Lattimer:1974slx,Eichler:1989ve,Freiburghaus1999a,Cowan:2019pkx}. All these facts imply that BNS mergers are unmissable research subjects from an astronomical point of view. They are also among the unique systems in the universe in which the most extreme (strongly self-gravitating, high-density, and high-temperature) environments in the universe are realized. Hence, the multi-messenger observation of BNS mergers is also an indispensable tool to extend our knowledge of fundamental physics.

Quantitative prediction of the merger dynamics and outcomes is crucial to correctly interpret the observed signals. Since the first simultaneous detection of gravitational waves and electromagnetic (EM) signals from a BNS (GW170817/AT2017gfo;~\citealt{TheLIGOScientific:2017qsa}), remarkable progress has been achieved in the theoretical understanding, particularly, in the studies based on numerical simulations. For example, recent numerical studies revealed the quantitative nature of mass ejection from BNS mergers, for which the processes can be broadly divided into two phases: At the onset of the merger, a fraction of neutron-rich matter is ejected by tidal force and collisional shock heating~\cite[e.g.,][]{Rosswog:1998hy,Ruffert:2001gf,Hotokezaka:2012ze}. After the merger, a massive neutron star (MNS) or a black hole (BH) surrounded by a strongly magnetized hot and dense accretion torus is formed~\citep[e.g.,][]{Price:2006fi,Kiuchi:2017zzg,Kiuchi:2022nin}. The magnetized central objects and accretion tori are considered to launch relativistic jets and outflows by magnetic pressure and tension, viscous heating due to magneto-hydrodynamical turbulence, and neutrino irradiation. Quantitative properties of the ejecta and the nucleosynthetic element abundances for each phase are studied by various groups together with their dependence on binary parameters, such as NS masses and NS equations of state~(EoS, \citealt{Hotokezaka:2012ze,Bauswein:2013yna,Wanajo:2014wha,Sekiguchi:2015dma,Foucart:2015gaa,Sekiguchi:2016bjd,Radice:2016dwd,Dietrich:2016hky,Bovard:2017mvn,Kiuchi:2017zzg,Dessart:2008zd,Metzger:2014ila,Perego:2014fma,Just:2014fka,Wu:2016pnw,Siegel:2017nub,Shibata:2017xdx,Lippuner:2017bfm,Fujibayashi:2017puw,Siegel:2017jug,Ruiz:2018wah,Fernandez:2018kax,Christie:2019lim,Perego:2019adq,Miller:2019dpt,Fujibayashi:2020qda,Fujibayashi:2020jfr,Fujibayashi:2020dvr,Bernuzzi:2020txg,Ciolfi:2020wfx,Vsevolod:2020pak,Foucart:2020qjb,Fernandez:2020oow,Mosta:2020hlh,Shibata:2021bbj,Shibata:2021xmo,Curtis:2021guz,Fujibayashi:2022ftg,Kiuchi:2022nin,Foucart:2022kon,Just:2023wtj,Curtis:2023zfo}; see~\citealt{Shibata:2019wef} for a review). The light curve modeling of EM counterparts, particularly for kilonovae, are also developed in this decade by employing numerical-simulation-based/motivated ejecta profiles and by performing radiative transfer simulations with realistic heating rates and/or detailed opacity tables~\citep[e.g.,][]{Kasen:2013xka,Kasen:2014toa,Barnes:2016umi,Wollaeger:2017ahm,Tanaka:2017lxb,Wu:2018mvg,Kawaguchi:2018ptg,Hotokezaka:2019uwo,Kawaguchi:2019nju,Korobkin:2020spe,Bulla:2020jjr,Zhu:2020eyk,Barnes:2020nfi,Nativi:2020moj,Kawaguchi:2020vbf,Wu:2021ibi,Just:2021vzy,Just:2023wtj}. 

However, there are still various open questions remaining. For example, whether the remnant NS has gravitationally collapsed into a BH or not is still being an open question for GW170817 due to the lack of the detection of post-merger gravitational waves in GW170817~\citep{LIGOScientific:2017fdd}. Such information is important, because it is connected to the underlying physics of the uncomprehended NS EoS~\citep[e.g.,][]{Margalit:2017dij,Rezzolla:2017aly,Shibata:2019ctb}. While we expect that the observation of the EM counterparts can provide a great hint to address this issue, it is still unclear from what observational features we can know about the fate of the remnant. Focusing particularly on the kilonova emission, a general consensus has not been yet reached for the property and origin of the ejecta in GW170817~\cite[e.g.,][]{Kasliwal:2017ngb,Cowperthwaite:2017dyu,Kasen:2017sxr,Villar:2017wcc,Waxman:2017sqv,Kawaguchi:2018ptg,Kawaguchi:2019nju,Bulla:2019muo,Almualla:2021znj,Kedia:2022onl,Bulla:2022mwo}. Determination of the ejecta property is crucial for understanding the post-merger evolution of the system and whether BNS mergers could be the major production site of $r$-process elements in the universe. 

To address these questions, quantitative understanding of the relation between the initial condition and/or underlying physics, and EM signals is important. For this purpose, conducting a study based on numerical simulations consistently starting from the merger to the phase of EM emission is a useful approach to link the observables that should be related to each other. In particular, for the kilonova modeling, it is important to accurately determine the ejecta profile for the rest-mass density and compositions at the time of kilonova emission ($>0.1$ d). Previous studies showed that the ejecta profile induces significant spacial dependence in radioactive heating as well as strong geometrical effects in radiative transfer, which have great impact on the resultant light curves~\citep{Kasen:2014toa,Wollaeger:2017ahm,Kawaguchi:2019nju,Bulla:2019muo,Zhu:2020inc,Darbha:2020lhz,Korobkin:2020spe,Almualla:2021znj,Kedia:2022onl}. However, there are still limited number of studies which provide the end-to-end modeling from the merger to observational outputs following the hydrodynamics evolution of all the ejecta components up to the time of kilonova emission (\citet{Kawaguchi:2020vbf,Kawaguchi:2022bub,Just:2023wtj}; see, however, \citet{Rosswog:2013kqa,Grossman:2013lqa,Collins:2022ocl,Neuweiler:2022eum} for the studies focusing on the dynamical ejecta components, and~\citet{Fernandez:2014bra,Fernandez:2016sbf,Foucart:2021ikp} in the context of BH-NS mergers). Given the situation that a number of BNS mergers will be observed in the next decades, the EM counterpart prediction based on the consistent simulations by taking the BNS diversity into account is an urgent task for correctly interpreting the observed data.

In this paper, we study the kilonova light curves of BNS mergers for the case that a remnant MNS forms and subsequently collapses to a BH within $20$\,ms after the onset of the merger (which we refer to as ``a short-lived case") consistently employing numerical-relativity (NR) results of~\citet{Kiuchi:2022ubj,Fujibayashi:2022ftg}. This paper is organized as follows: In Section~\ref{sec:method}, we describe the method employed in this study. In Section~\ref{sec:model}, we describe the BNS models we study in this work. In Section~\ref{sec:result}, we present the property of the ejecta obtained by the long-term hydrodynamics evolution and the kilonova light curves obtained by radiative-transfer simulations. Finally, we discuss the implication of this paper in Section~\ref{sec:discuss}. Throughout this paper, $c$ denotes the speed of light.


\section{Method}~\label{sec:method}
Merger ejecta of a BNS are expected to be homologously expanding at the time of kilonova emission ($\gtrsim 0.1\,{\rm d}$). To obtain the ejecta profile in the homologously expanding phase, we follow the same procedures as in the previous work~\citep{Kawaguchi:2020vbf,Kawaguchi:2022bub}; adopting the outflow data obtained by NR simulations as the inner boundary condition~\citep{Fujibayashi:2022ftg}, the hydrodynamics evolution of merger ejecta is calculated by employing an axisymmetric relativistic hydrodynamics code developed in~\citet{Kawaguchi:2020vbf,Kawaguchi:2022bub}. In the following, to distinguish between the present simulation and NR simulation, we refer to the present hydrodynamics simulations as the HD simulations.

In the hydrodynamics code, relativistic hydrodynamics equations in the spherical coordinates are solved taking into account the effect of fixed-background gravity of a non-rotating BH metric in the isotropic coordinates. Radioactive-decay heating of heavy elements is also taken into account by referring to the nucleosynthesis results computed for each ejecta fluid element in the NR simulation (see~\citet{Fujibayashi:2022ftg} for the details). We employ the ideal-gas EoS with the adiabatic index of $\Gamma=4/3$. For the HD simulations, the uniform grid spacing with $N_\theta$ grid points is prepared for the polar angle $\theta$, while for the radial direction, the following non-uniform grid structure is employed; the $j$-th radial grid point is given by
\begin{align}
	{\rm ln}\,r_j={\rm ln}\left(\frac{r_{\rm out}}{r_{\rm in}}\right)\frac{j-1}{N_r}+{\rm ln}\,r_{\rm in},\,j=1\cdots N_r+1,\label{eq:grid}
\end{align} 
where $r_{\rm in}$ and $r_{\rm out}$ denote the inner and outer radii of the computational domain, respectively, and $N_r$ denotes the total number of the radial grid points. In the present work, we employ $(N_r,N_\theta)=(2048,256)$, and $r_{\rm in}$ and $r_{\rm ext}$ are initially set to be $8,000\,{\rm km}$ and $10^3\,r_{\rm in}$, respectively. We employ the same time origin for the HD simulations as in the NR simulations for the post-merger evolution.

To import the outflow data from the NR simulations of~\citet{Fujibayashi:2022ftg} to the present HD simulations, the time-sequential hydrodynamics property of the outflow is extracted at $r=r_{\rm in}$ in the NR simulations, and is used as the boundary condition at the inner radius, $r=r_{\rm in}$, of the HD simulations. The NR simulation data are run out at $t> 5$\,s, and after then, the HD simulation is continued by setting a very small floor-value, which is negligible for the ejecta dynamics, to the rest-mass density of the inner boundary. To follow the evolution of ejecta even after the high velocity edge of the outflow reaches the outer boundary of our HD simulation, the radial grid points are added to the outside of the original outer boundary, while at the same time the innermost radial grid points are removed so as to keep the total number of the radial grid points. By this prescription, the value of $r_{\rm in}$ is increased in the late phase of the HD simulations. The outermost radial grids are added so that the location of the outer radial boundary, $r_{\rm out}$, is always $10^3 r_{\rm in}$. We note that the total mass lost by removing the inner radial grids is always much smaller ($\lesssim 10^{-4}\,M_\odot$) than the post-merger ejecta mass.

The light curves of kilonovae are calculated using a wavelength-dependent radiative transfer simulation code~\citep{Tanaka:2013ana,Tanaka:2017qxj,Tanaka:2017lxb,Kawaguchi:2019nju,Kawaguchi:2020vbf}. In this code, the photon transfer is simulated by a Monte Carlo method for given ejecta profiles composed of the density, velocity, and element abundance under the assumption of the homologous expansion. The time-dependent thermalization efficiency is taken into account following an analytic formula derived by~\citet{Barnes:2016umi}. The ionization and excitation states are determined under the assumption of the local thermodynamic equilibrium (LTE) by using the Saha's ionization and Boltzmann excitation equations. The impact of this assumption will be discussed in Appendix~\ref{app:nonLTE}.

For the photon-matter interaction, bound-bound, bound-free, and free-free transitions and electron scattering are taken into account for the transfer of optical and infrared photons~\citep{Tanaka:2013ana,Tanaka:2017qxj,Tanaka:2017lxb}. The formalism of the expansion opacity~\citep{1983ApJ...272..259F,1993ApJ...412..731E,Kasen:2006ce} and the new line list derived in~\citet{Domoto:2022cqp} are employed for the bound-bound transitions. In this line list, the atomic data of VALD~\citep{1995A&AS..112..525P,1999A&AS..138..119K,2015PhyS...90e4005R} or Kurucz's database~\citep{1995all..book.....K} are used for $Z=20$–$29$, while the results of atomic calculations from~\citet{Tanaka:2019iqp} are used for $Z=30$--$88$. For Sr II, Y I, Y II, Zr I, Zr II, Ba II, La III, and Ce III, which are the ions producing strong lines, line data are replaced with those calibrated with the atomic data of VALD and NIST database~\citep{NIST}.

The radiative transfer simulations are performed from $t=0.1\,{\rm d}$ to $30\,{\rm d}$ employing the density and internal energy profiles of the HD simulations at $t=0.1\,{\rm d}$. The spatial distributions of the heating rate and element abundances are determined by the table obtained by the nucleosynthesis calculations referring to the injected time and angle of the fluid elements. Note that the element abundances at $t=1\,{\rm d}$ are used during the entire time evolution in the radiative transfer simulations to reduce the computational cost, but this simplified prescription gives an only minor systematic error on the resultant light curves as illustrated in~\citet{Kawaguchi:2020vbf}.

\section{Model}~\label{sec:model}
\begin{table*}
\caption{Key model parameters. The columns describe the model name, the EoS adopted, the masses of the NSs, type of the MNS evolution, the ejecta mass evaluated in the NR simulations  ($M^{\rm NR}_{\rm eje}$, $M^{\rm NR}_{\rm dyn}$, and  $M^{\rm NR}_{\rm post}$ denote the total, dynamical, and post-merger masses, respectively; see~\citealt{Fujibayashi:2020dvr,Shibata:2021xmo,Fujibayashi:2022ftg}), and the ejecta mass evaluated in the HD simulations at $t=0.1\, {\rm d}$, $M^{\rm HD}_{\rm eje}$, respectively. ``short-lived'', ``long-lived'', and ``long-lived with strong dynamo'' denote the cases for which the remnant MNS collapses to a BH within 20 ms, survives for $\gg$ 1 s, and survives for $\gg$ 1 s with significant magnetic dynamo effects, respectively. The values for $M^{\rm NR}_{\rm eje}$ are calculated by integrating the mass flux at the sphere with radius 8,000\,km over time in 2D NR simulations. We then subtract from $M^{\rm NR}_{\rm eje}$ the mass of dynamical ejecta $M^{\rm NR}_{\rm dyn}$, which is evaluated in the corresponding 3D NR simulations with the Bernoulli criterion to obtain the contribution of the post-merger ejecta $M^{\rm NR}_{\rm post}$.}
\centering
\begin{tabular}{c|c|c|c|c|c}\hline
Model	& EoS &	$(m_1\,[M_\odot],m_2\,[M_\odot])$&  MNS evolution  &	$M^{\rm NR}_{\rm eje}(M^{\rm NR}_{\rm dyn},M^{\rm NR}_{\rm post})\,[10^{-2}M_\odot]$&	$M^{\rm HD}_{\rm eje}\,[10^{-2}M_\odot]$\\\hline\hline
SFHo-135135	&	SFHo &	$(1.35,1.35)$&short-lived&	$1.0\,(0.73,\,0.25)$&	$1.0$\\
SFHo-130140	&	SFHo &	$(1.30,1.40)$&short-lived&	$1.0\,(0.48,\,0.50)$&	$0.9$\\
SFHo-125145	&	SFHo &	$(1.25,1.45)$&short-lived&	$1.2\,(0.64,\,0.60)$&	$1.1$\\
SFHo-120150	&	SFHo &	$(1.20,1.50)$&short-lived&	$1.6\,(0.45,\,1.1)$&	$1.5$\\
SFHo-125155	&	SFHo &	$(1.25,1.55)$&short-lived&	$1.5\,(0.95,\,0.55)$&	$1.4$\\\hline
DD2-135135	    &	DD2&	$(1.35,1.35)$&long-lived&	$7.6\,(0.15,\,7.5)$&	$6.5$\\
MNS75a	        &	DD2&	$(1.35,1.35)$&long-lived with strong dynamo&	$9.4\,(0.15,\,9.3)$&	$8.4$\\\hline
\end{tabular}
\label{tb:model}
\end{table*}

In this work, we employ the NR outflow profiles obtained in~\citet{Fujibayashi:2022ftg} as the input for the HD simulations. The key quantities of each model are summarized in Table~\ref{tb:model}. The first four models listed in Table~\ref{tb:model} are BNSs with the total gravitational mass (at the infinite separation) of $2.7\,M_\odot$ but with various mass ratios in the range of 0.8--1.0. We also study an unequal mass BNS with a larger total gravitational mass ($2.8\,M_\odot$), which we refer to as SFHo-125155. The SFHo EoS~\citep{Steiner:2012rk} supplemented by the Timmes (Helmholtz) EoS~\citep{2000ApJS..126..501T} for the low density part is employed. For all the models employing the SFHo EoS, a remnant MNS is formed after the merger, but it collapses to a BH within $\approx20\,{\rm ms}$. We note that these mass ranges of the BNSs with a short-lived remnant broadly cover the range of the mass estimation obtained by the gravitational-wave data analysis of GW170817~\citep{LIGOScientific:2017vwq,LIGOScientific:2018hze}. The BNS models which result in the formation of an MNS surviving for a long time $(>1\,{\rm s};$~\citealt{Fujibayashi:2020dvr,Shibata:2021xmo}) are also shown in Table~\ref{tb:model} for comparison purposes (see also ~\citealt{Kawaguchi:2020vbf,Kawaguchi:2022bub}). 
The NR simulations are performed by a general-relativistic viscous neutrino-radiation hydrodynamics code with the dimensionless alpha viscous parameter of $\alpha=0.04$ ~\citep{Fujibayashi:2020dvr,Fujibayashi:2022ftg} except for MNS75a in which general-relativistic neutrino-radiation resistive-magnetohydrodynamics code is employed to take the magnetic dynamo effects into account~\citep{Shibata:2021xmo}.



The ejecta mass evaluated in the NR simulations is also listed in Table~\ref{tb:model}. The total ejecta mass increases as the mass ratio of the BNS deviates from unity due to the increase in the torus mass, and hence, the ejecta mass of the post-merger component. Broadly speaking, the mass of the dynamical ejecta tends to decrease as the binary becomes more asymmetric (but not so monotonically). This reflects the fact that, for an asymmetric binary, the tidal-interaction-driven component dominates the dynamical ejecta rather than the collisional shock-driven component, of which the launching mechanism is more efficient in mass ejection than the former. The total ejecta mass of the BNS merger for which the remnant MNS collapses to a BH in a short time is an order of magnitude smaller than that for the BNS which results in the formation of an MNS surviving for a long time $(>1\,{\rm s};$~\citealt{Fujibayashi:2020dvr,Shibata:2021xmo}). Note that for the latter case, the total ejecta mass is dominated by the post-merger ejecta.

\section{Results}~\label{sec:result}

\subsection{Ejecta profiles}
For all the models, we find that the total internal energy of ejecta is smaller by $\approx4$ order of magnitudes than the total kinetic energy at $t=0.1\,{\rm d}$ and that the mass-averaged deviation of the velocity field from that in the later homologous expanding phase ($v^r=r/t$ with $v^r$ being the radial velocity) is as small as $10^{-3}$ at $t=0.1\,{\rm d}$. This shows that the homologous expansion is well achieved for $t \geq 0.1\,{\rm d}$.

The total mass in the computational domain measured at $t=0.1\,{\rm d}$, $M^{\rm HD}_{\rm eje}$, is listed in Table~\ref{tb:model}. Note that the matter is in the homologously expanding phase at $t=0.1\,{\rm d}$, and hence, $M^{\rm HD}_{\rm eje}$ can be regarded as the total ejecta mass. 
It is found that $M^{\rm HD}_{\rm eje}$ is slightly smaller than $M^{\rm NR}_{\rm eje}$ for some of the models. This is a consequence of the fact that a fraction of the matter falls back across the inner boundary as the pressure support from the inner boundary vanishes when the outflow data run out. While a fraction of the matter can actually experience such fall-back due to the deceleration by the pressure from the precedingly ejected matter, our treatment of suddenly vanishing pressure support on the inner boundary at the run-out time of NR data may artificially increase the mass of the fall-back matter. Nevertheless, as found in our previous studies~\citep{Kawaguchi:2020vbf,Kawaguchi:2022bub}, the contribution of such marginally unbound matter to the kilonova emission is minor because it has only low velocity and has only a small contribution to the emission due to the long diffusion time scale.

\begin{figure*}
 	 \includegraphics[width=.49\linewidth]{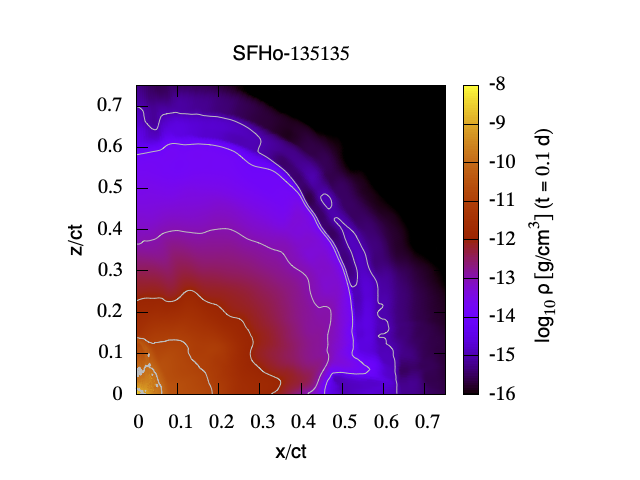}
 	 \includegraphics[width=.49\linewidth]{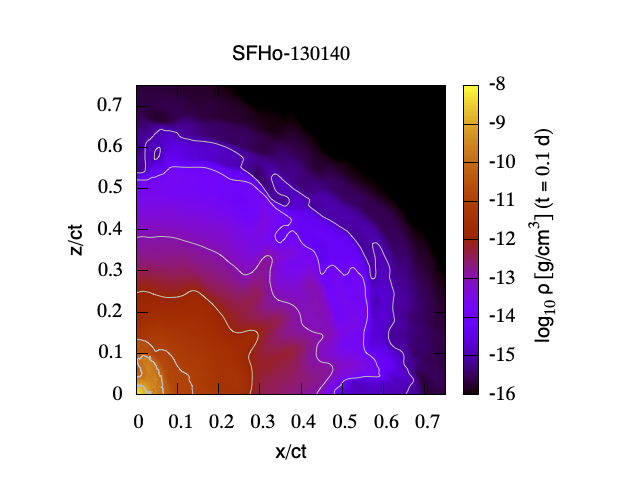}\\
 	 \includegraphics[width=.49\linewidth]{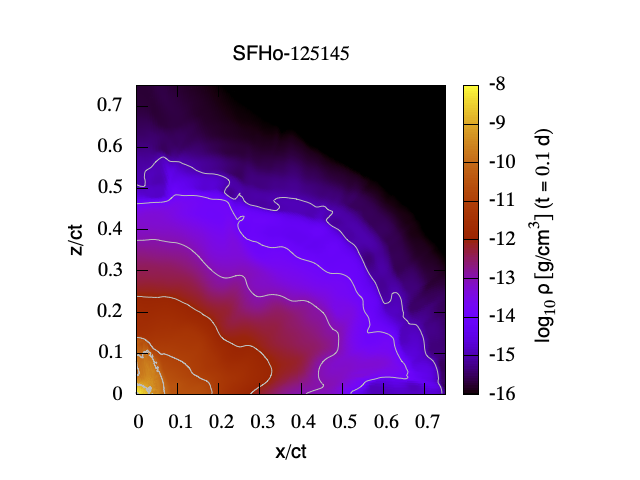}
 	 \includegraphics[width=.49\linewidth]{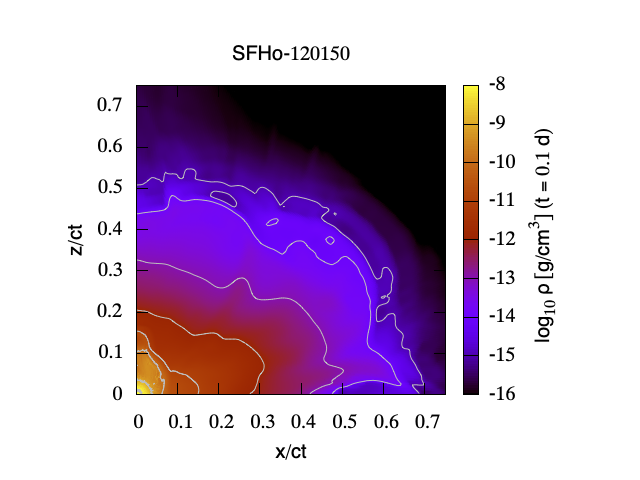}
 	 \caption{Rest-mass density profiles at $t=0.1\,{\rm d}$ obtained by the HD simulations. The top-left, top-right, bottom-left, and bottom-right panels display the results for models SFHo-135135, SFHo-130140, SFHo-125145, and SFHo-120150, respectively. The gray curves in each panel denote the contour lines of $10^{-15}$, $10^{-14}$, $10^{-13}$, $10^{-12}$, $10^{-11}$, $10^{-10}$, and $10^{-9}\,{\rm g/cm^3}$ from outside.}
	 \label{fig:rho_prof}
\end{figure*}

\begin{figure*}
 	 \includegraphics[width=.49\linewidth]{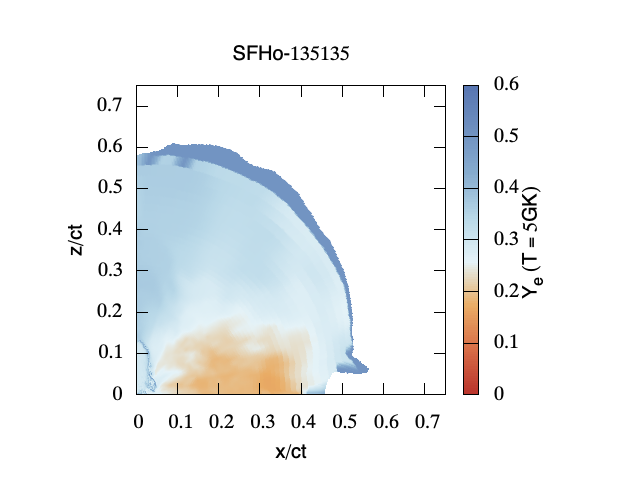}
 	 \includegraphics[width=.49\linewidth]{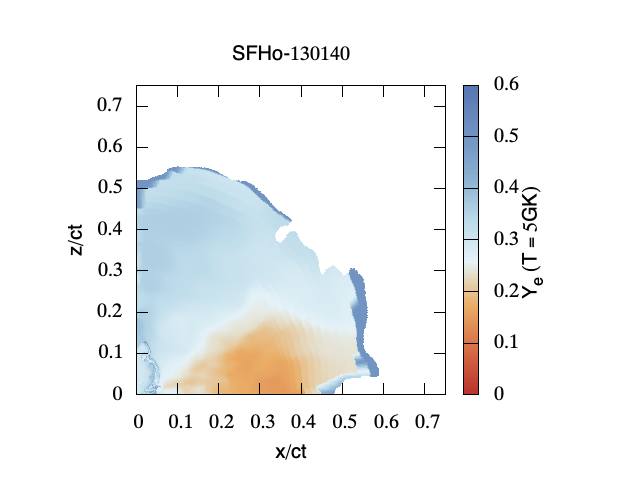}\\
 	 \includegraphics[width=.49\linewidth]{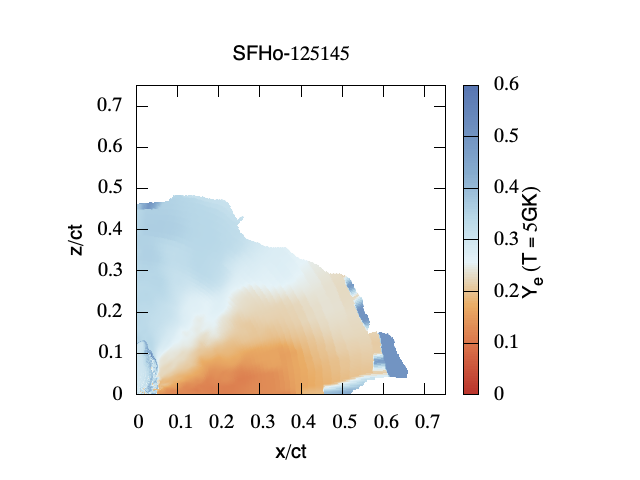}
 	 \includegraphics[width=.49\linewidth]{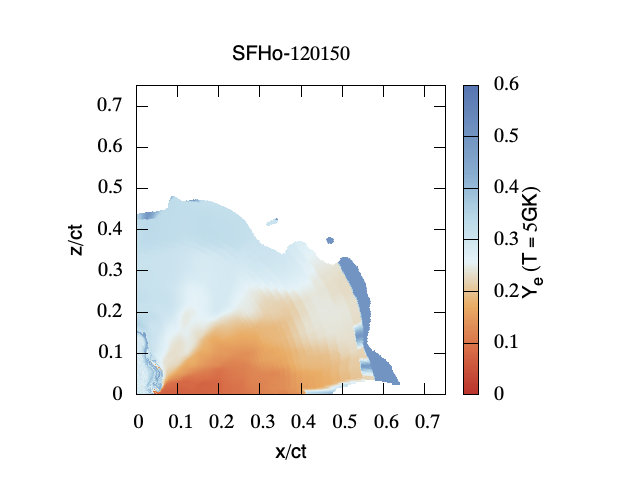}
 	 \caption{The same as Fig.~\ref{fig:rho_prof} but for the electron fraction, $Y_e$. The value of $Y_e$ is evaluated when the temperature of the fluid element decreases to $T=5\,{\rm GK}$. Note that only the region of which the rest-mass density at $t=0.1\,{\rm d}$ is higher than $10^{-14}\,{\rm g/cm^3}$ is shown.}
	 \label{fig:ye_prof}
\end{figure*}

\begin{figure*}
 	 \includegraphics[width=.49\linewidth]{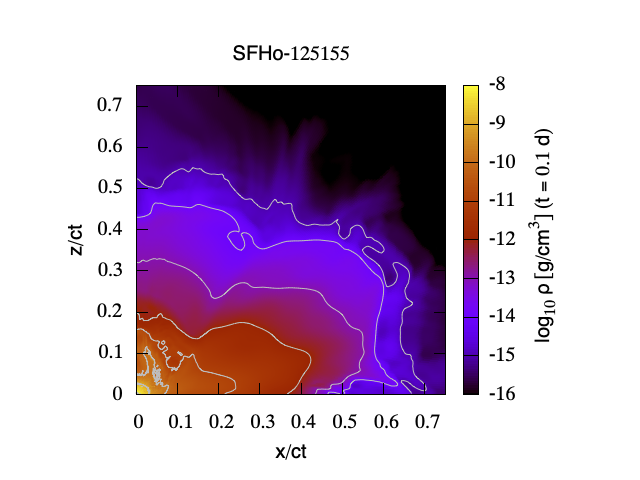}
 	 \includegraphics[width=.49\linewidth]{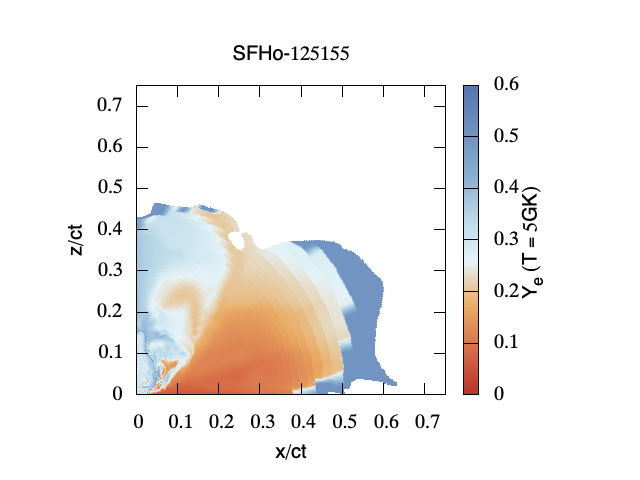}
 	 \caption{The same as Figures~\ref{fig:rho_prof} and ~\ref{fig:ye_prof} but for SFHo-125155.}
	 \label{fig:prof_125155}
\end{figure*}

First, we focus on the BNS models of which the total mass is $2.7\,M_\odot$ to see the effect of the binary mass ratio. Fig.~\ref{fig:rho_prof} shows the rest-mass density profiles at $t=0.1\,{\rm d}$ obtained by the HD simulations for models SFHo-135135, SFHo-130140, SFHo-125145, and SFHo-120150. The dynamical ejecta component located at $x/ct\gtrsim 0.05$ or $z/ct\gtrsim 0.15$ exhibits a broadly spherical morphology in the rest-mass density structure. On the other hand, the post-merger ejecta component, which is present in $x/ct\lesssim 0.05$ and $z/ct\lesssim 0.15$, exhibits a mildly prolate shape (see Fig.~\ref{fig:ye_prof} for a clearer distinction between the dynamical and post-merger ejecta components). These characteristics of the density profile are in broad agreement with the ejecta profile obtained in our previous studies~\citep{Kawaguchi:2020vbf,Kawaguchi:2022bub}, in which BNSs result in long-lived MNSs (with the lifetime of $>1\,{\rm s}$).

Taking a closer look, the dynamical ejecta show a relatively more prolate shape for an equal-mass BNS (SFHo-135135), while relatively more oblate shapes are seen for unequal mass cases (SFHo-125145 and SFHo-120150). This reflects the fact that the tidally driven component which spreads preferentially toward the equatorial direction dominates in the dynamical ejecta for an asymmetric binary over the collisional-shock-driven component which spreads in a more spherical manner. 

Fig.~\ref{fig:ye_prof} shows the electron fraction ($Y_e$) profiles at $t=0.1\,{\rm d}$ for models SFHo-135135, SFHo-130140, SFHo-125145, and SFHo-120150. Here, the value of $Y_e$ is evaluated when the temperature of the fluid element decreases to $T=5\,{\rm GK}$ ($=5\times10^9$\,K). 
A clear boundary-like feature starting from $x/ct\approx 0.05$ on the equatorial plane to $z/ct\lesssim 0.15$ along the polar axis is seen for all the models. This corresponds to the boundary between the dynamical and post-merger ejecta components. The dynamical ejecta has a clear angular dependence in the $Y_e$ profile. With $\theta$ being the angle measured from the polar axis, the value of $Y_e$ of the dynamical ejecta is higher than $0.3$ for $\theta\lesssim 45^\circ$--$60^\circ$, while it is lower than $0.3$ for $\theta\gtrsim 45^\circ$--$60^\circ$. This clearly reflects the difference in the mass ejection mechanism; the former is shock-heating-driven and the latter is tidally driven. The dynamical ejecta for unequal mass BNSs have relatively more extended distribution and lower $Y_e$ values along the equatorial direction than those for the equal-mass case. This also reflects the fact that the tidally driven component dominates the dynamical ejecta and the ejecta experience a relatively small rise in temperature resulting from the shock heating for the unequal mass cases. On the other hand, the post-merger ejecta has only weak angular dependence in the $Y_e$ value, which is always $\gtrsim0.3$. These profiles of $Y_e$ are also in broad agreement with the previous results of BNS mergers that result in long-lived remnant MNSs~\citep{Kawaguchi:2020vbf,Kawaguchi:2022bub} and the results of BNS mergers in which the remnant survives for a moderately long time (0.1--1 s) ~\citep{Just:2023wtj}.

Fig.~\ref{fig:prof_125155} shows the rest-mass density and electron fraction profiles for model SFHo-125155. The qualitative features of the rest-mass density and $Y_e$ profiles for this model are the same as those for other models with the total mass of 2.7 $M_\odot$, but the oblate shape and low-$Y_e$ value region are more pronounced than those for the models shown in Figs.~\ref{fig:rho_prof} and \ref{fig:ye_prof}. This reflects the fact that SFHo-125155 has the largest dynamical ejecta mass dominated by the  tidally driven component as the consequence of the large asymmetry in the NS masses. 

Figs.~\ref{fig:rho_prof}--\ref{fig:prof_125155} illustrate that the profiles of the rest-mass density and electron fraction depend sensitively on the total mass and mass ratio of the binaries. In the following we will show that the light curve and the spectral evolution depend on these differences, although the type of the remnant (either a short-lived or long-lived neutron star is formed as a remnant) has more impact on the brightness of the kilonova light curve.

\subsection{Kilonova light curves}
\begin{figure*}
 	 \includegraphics[width=.48\linewidth]{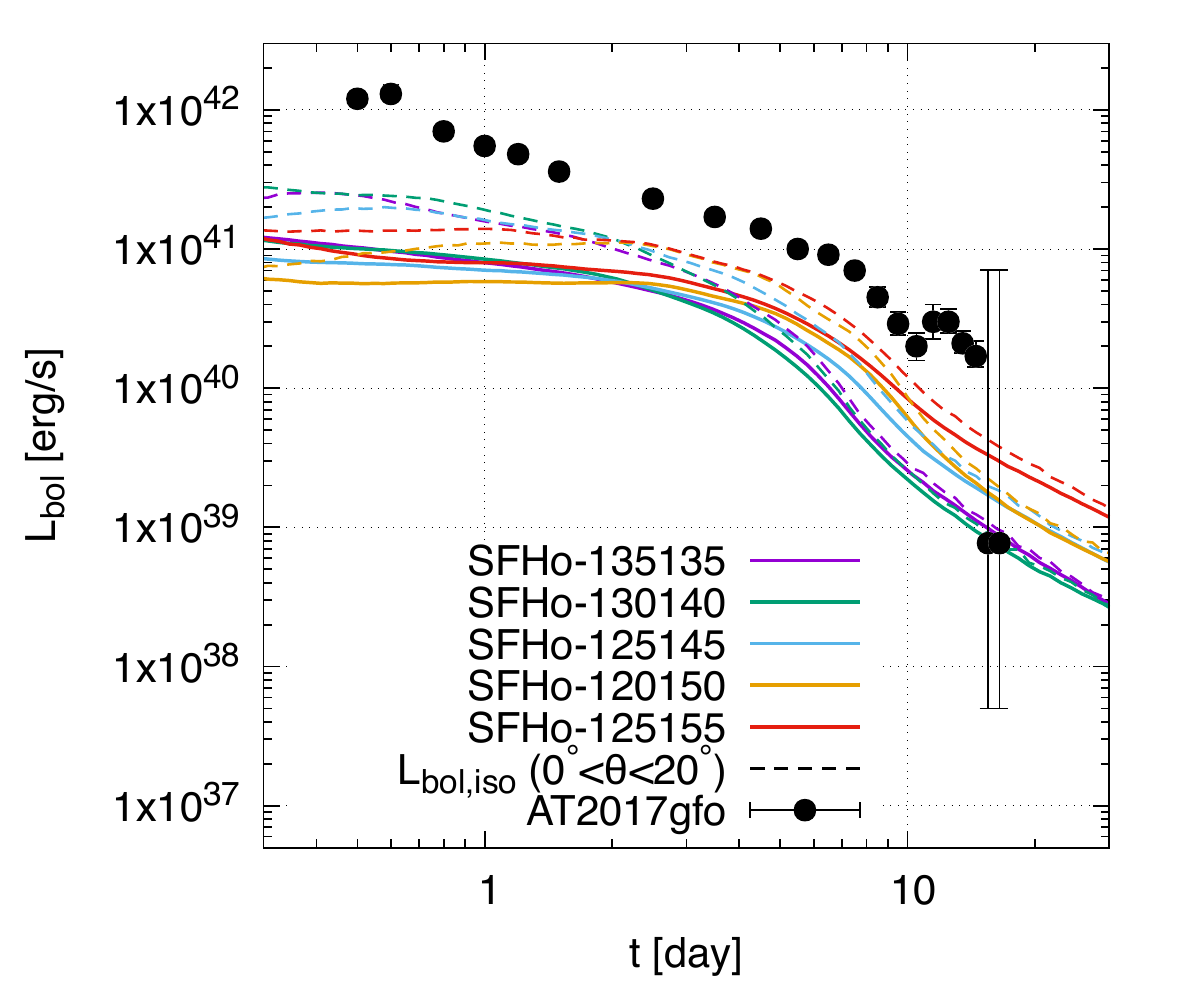}
 	 \includegraphics[width=.48\linewidth]{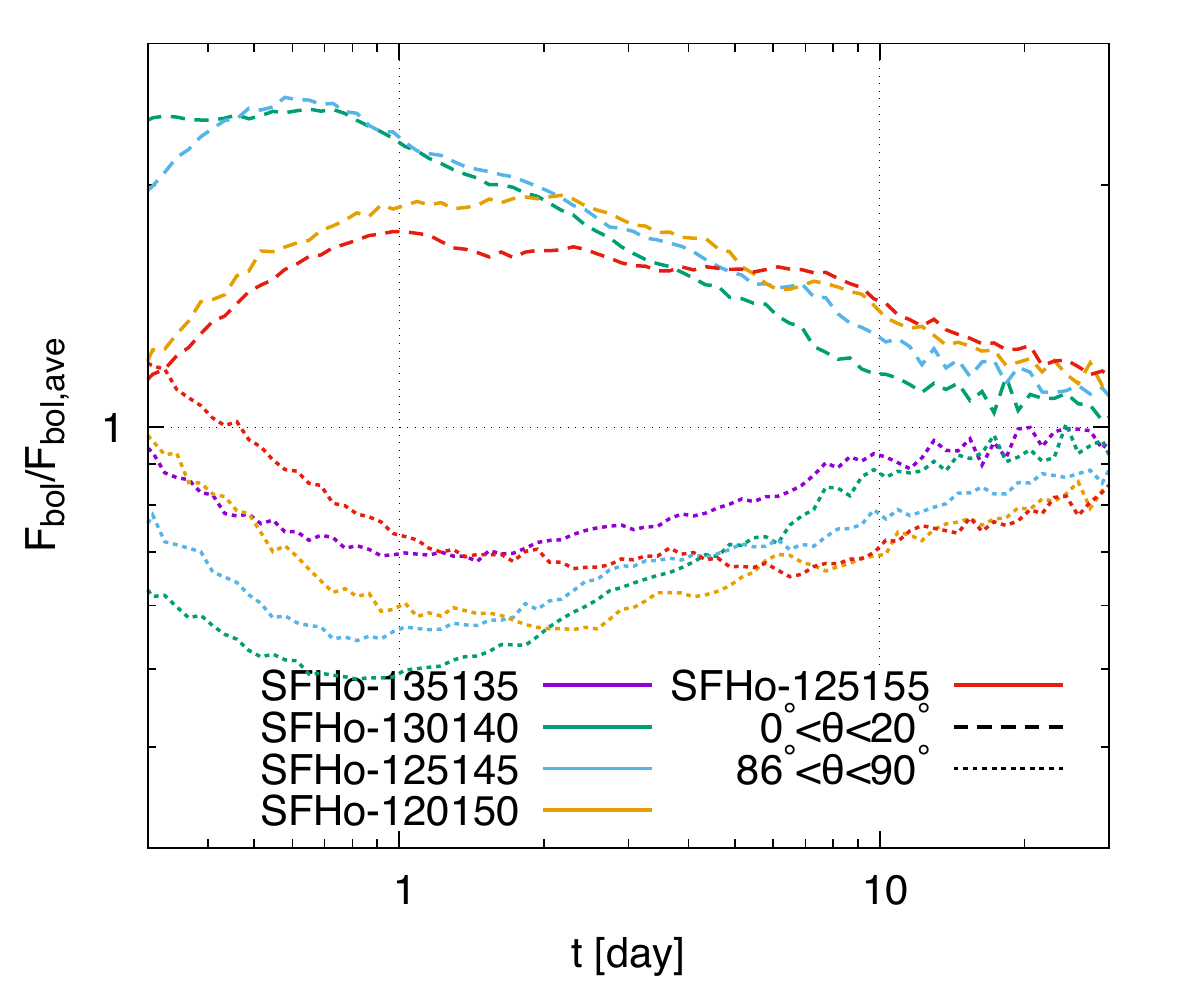}
 	 \caption{(Left panel) The bolometric light curves for all the models considered in this paper. The solid and dashed curves denote the total and isotropically equivalent bolometric luminosities (the latter measured from the polar direction, $0^\circ\le\theta\le20^\circ$), respectively. The isotropically equivalent bolometric luminosity observed in AT2017gfo is shown by the filled circles adopting the data in~\citet{Waxman:2017sqv} with the distance of 40 Mpc. (Right panel)  the ratios of the bolometric fluxes measured from the polar (the dashed curves; $0^\circ\le\theta\le20^\circ$) and equatorial directions (the dotted curves; $86^\circ\le\theta\le90^\circ$) to those of spherical average.}
	 \label{fig:lbol}
\end{figure*}

The left panel of Fig.~\ref{fig:lbol} shows the results of the bolometric light curves obtained by radiative-transfer simulations. The solid and dashed curves denote, respectively, the total and isotropically equivalent bolometric luminosities (the latter measured from the polar direction, $0^\circ\le\theta\le20^\circ$). For all the models, the bolometric light curves show approximately flat features with the luminosity of  $\sim 10^{41}\,{\rm erg/s}$ for $0.3\,{\rm d}\le t\le$3--5 d, and decline rapidly after 3--5 d. As the ejecta mass increases, the epoch at which the bolometric light curve starts rapidly declining is delayed, and the luminosity after the decline becomes larger. 
This reflects the larger total optical depth and deposition energy for larger ejecta mass models.

The right panel of Fig.~\ref{fig:lbol} shows the ratios of the bolometric fluxes measured from the polar ($0^\circ\le\theta\le20^\circ$) and equatorial directions ($86^\circ\le\theta\le90^\circ$) to those of spherical average. The isotropically equivalent luminosities measured from the polar and equatorial directions are brighter and fainter by a factor of $\approx 2$, respectively, at $t\sim 1\,{\rm d}$ due to the preferential diffusion of photons in the presence of optically thick dynamical ejecta around the equatorial plane~\citep{Kawaguchi:2018ptg,Kawaguchi:2019nju}. However, such effects become less significant in the late phase ($\gtrsim 10\,{\rm d}$) as the optical depth of the ejecta decreases due to the expansion. The viewing-angle dependence of the bolometric light curves is sustained for a longer time scale as the binary becomes more asymmetric. This reflects the fact that the tidally driven component of dynamical ejecta has more mass and a lower value of $Y_e$ for more asymmetric binaries, resulting in more opaque ejecta.

None of the model light curves in the left panel of Fig.~\ref{fig:lbol} can explain the observed brightness of the kilonova associated with GW170817 (AT2017gfo). The bolometric light curves are always below the observational data from 0.5\,d to 17\,d except for the last two data points in the plot. This is the case even if the enhancement of the brightness due to geometrical effects is taken into account (see the dashed curves in the left panel of Fig.~\ref{fig:lbol}, which denote the light curves measured from the polar direction). This is primarily due to the smallness of the ejecta mass, which leads to insufficient total radioactive deposition energy to explain the observation of AT2017gfo. Our results indicate that a BNS for which a remnant MNS collapses to a BH in a short time ($t\lesssim 20\,{\rm ms}$) is unlikely to be the progenitor of GW170817. We note that our light curves are fainter than the results of~\citet{Just:2023wtj}, which considers the cases that a remnant MNS survives for a relatively longer time scale before it collapses to a BH (at $t=0.1$--1\,s after the onset of the merger). This simply reflects the fact that the total ejecta mass is smaller for our present models.

\begin{figure}
 	 \includegraphics[width=0.95\linewidth]{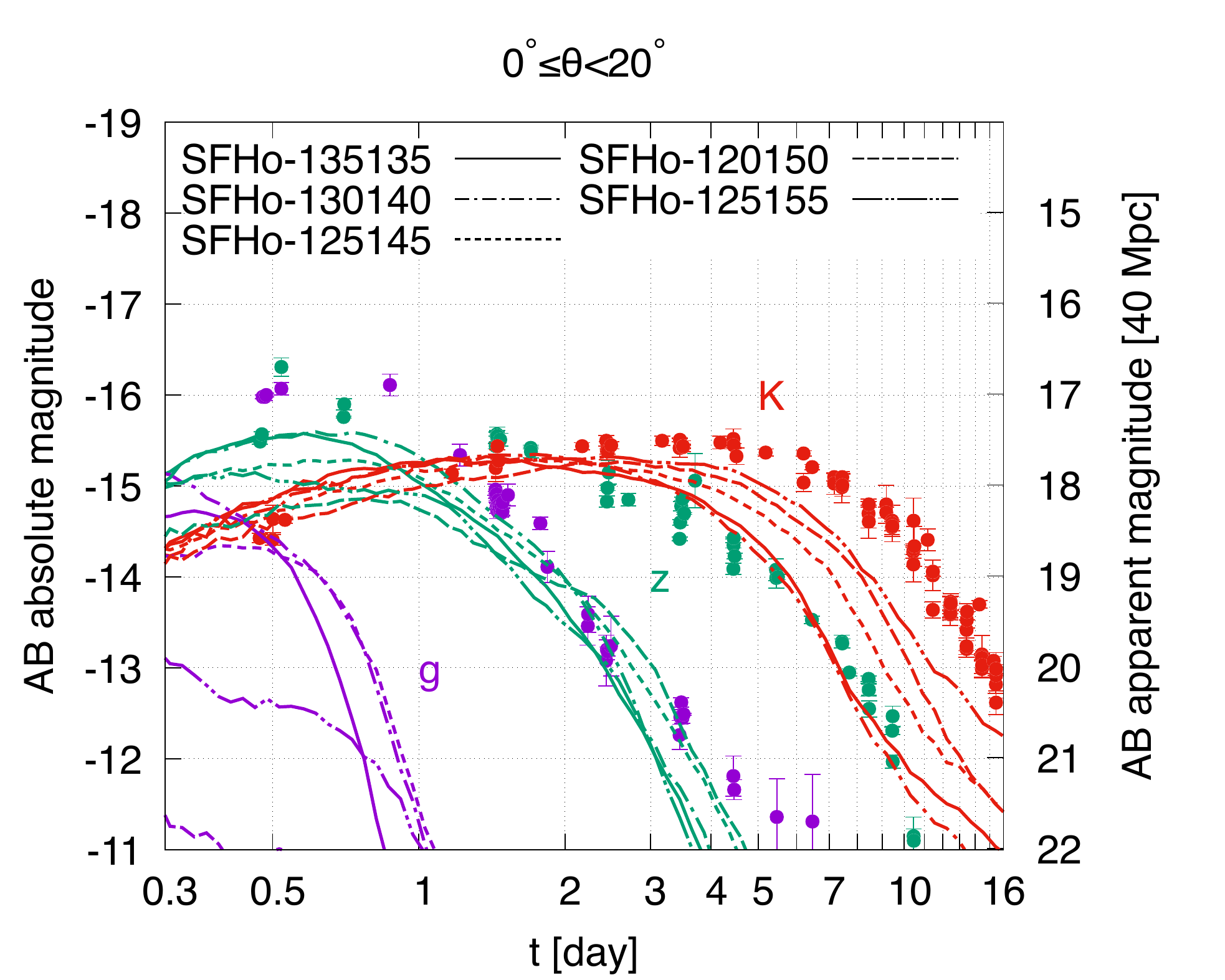}\\
 	 \includegraphics[width=0.95\linewidth]{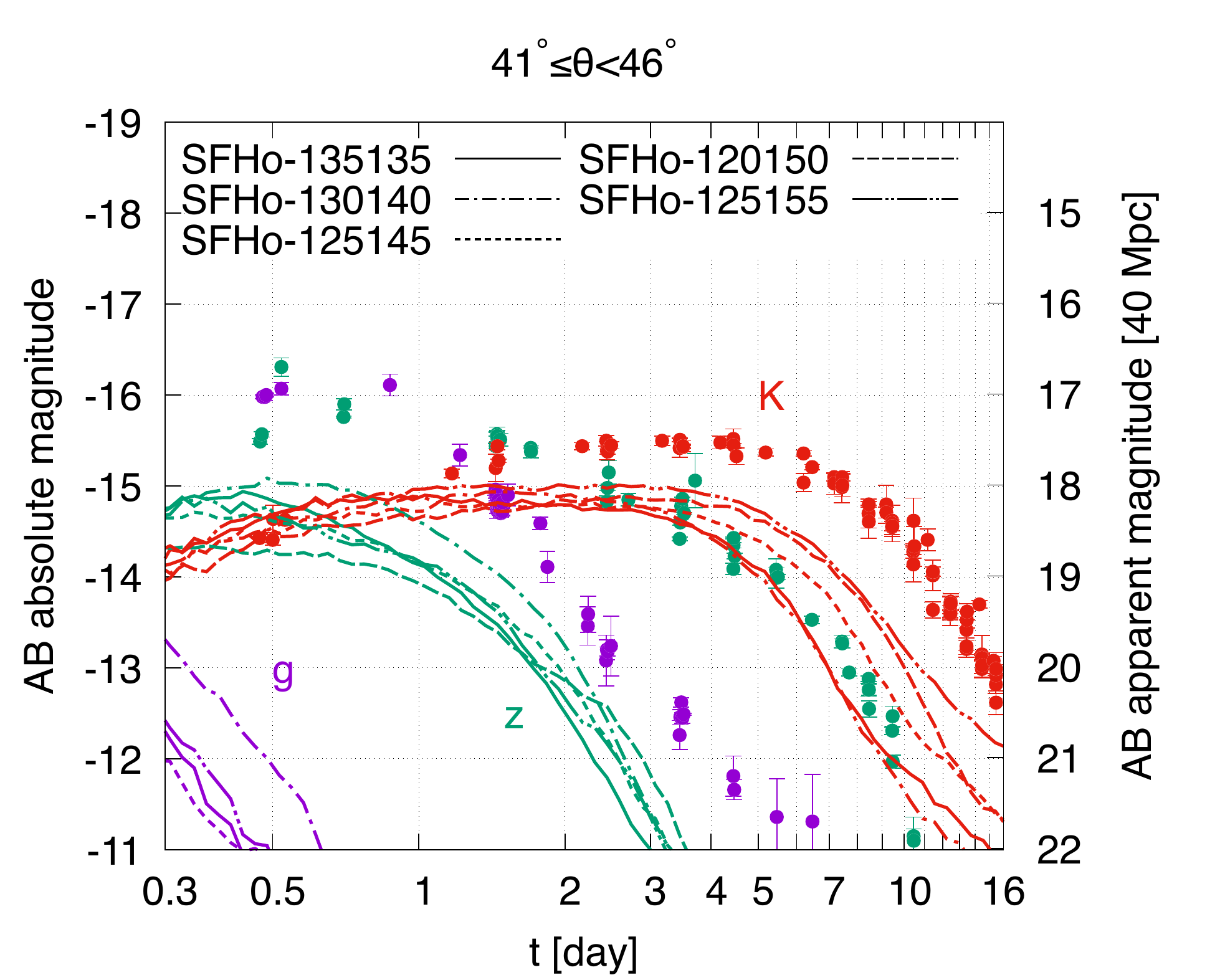}\\
 	 \includegraphics[width=0.95\linewidth]{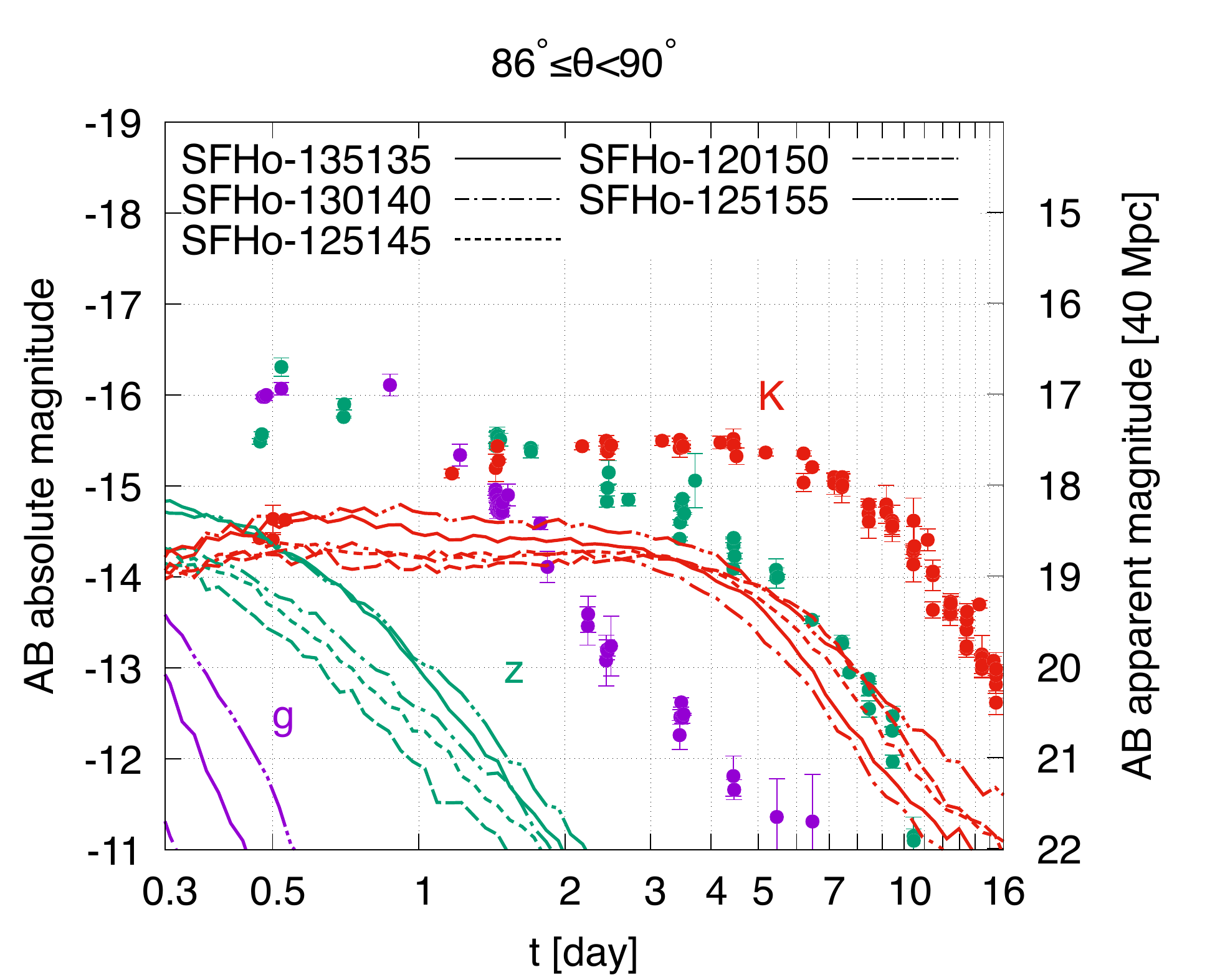}
 	 \caption{{\it gzK}-band light curves. The top, middle, and bottom panels denote the light curves observed from $0^\circ\le\theta\le20^\circ$, $41^\circ\le\theta\le46^\circ$, and $86^\circ\le\theta\le90^\circ$, respectively. The purple, green, and red curves denote the {\it g},  {\it z}, and  {\it K}-band light curves, respectively. The data points denote the observation data of AT2017gfo taken from~\citet{Villar:2017wcc} with the distance of 40 Mpc.}
	 \label{fig:mag}
\end{figure}

Fig.~\ref{fig:mag} shows the {\it gzK}-band light curves for all the models of the short-lived cases listed in Table~\ref{tb:model}. The obtained light curves show the broadly similar properties to those obtained by the observation of AT2017gfo as well as the previous studies for a kilonova with multiple ejecta components~\cite[e.g.,][]{Kasen:2014toa,Wollaeger:2017ahm,Kawaguchi:2018ptg,Bulla:2019muo}; the optical emission lasts for a short time scale ($\sim 1\,{\rm d}$), and the near-infrared (NIR) emission lasts for a longer time scale ($\sim 10\,{\rm d}$). The emission becomes faint as the viewing angle measured from the axis of symmetry increases. This primarily reflects the spatial dependence of element abundances (see Figs.~\ref{fig:ye_prof} and~\ref{fig:prof_125155}). The viewing-angle dependence is more pronounced for the emission in the optical wavelength (i.e., in the {\it g}-band) due to the so-called lanthanide-curtain effects in the presence of low-$Y_e$ dynamical ejecta around the equatorial plane~\citep{Kasen:2014toa,Wollaeger:2017ahm,Kawaguchi:2019nju,Bulla:2019muo,Zhu:2020inc,Darbha:2020lhz,Korobkin:2020spe}. 

Interestingly, the peak magnitudes in the NIR wavelengths (i.e., in the {\it K}-band) do not significantly differ among the models regardless of the difference in the ejecta mass. However, the time scale for the emission to sustain the brightness close to the peak becomes shorten as the total ejecta mass decreases. The light curves in the optical wavelengths observed from the polar direction also show similar shapes among the models except for the most asymmetric BNSs (SFHo-120150 and SFHo-125155) for which the {\it g}-band light curves are fainter by $\geq 1$ mag than those for the other models. We find that the strong suppression of the optical emission for the most asymmetric BNS models is due to the fact that the polar regions are more polluted by the lanthanide elements. The difference in the brightness of the optical emission observed from the equatorial direction among the models simply reflects the difference in the dynamical ejecta mass (see Table~\ref{tb:model}).

Our result implies that an earlier follow-up observation than in GW170817/AT2017gfo is needed to observe the kilonova emission in the optical band for the short-lived BNS formation. For example, for the hypothetical distance of 200 Mpc, the {\it g}-band emission can only be detected by the observation within 0.5--1 d with the sensitivity deeper than 22 mag, which requires telescopes larger than 2 m-classes~\citep{Nissanke:2012dj}. Also, such a detection can be achieved only for the case that the event is face-on, but we should note that it could be hidden by the GRB afterglow emission. In the {\it z} band, the emission lasts for a longer time scale, but yet the observation within 1 d is needed with 2 m and 4 m-class telescopes, respectively, to find kilonovae for the case of $\theta\leq 45^\circ$. The NIR follow-up observation by a telescope larger than 4-m classes, such as VISTA~\citep{Ackley:2020qkz}, can detect the kilonova emission up to 5 d after the onset of the merger with the hypothetical distance of 200 Mpc and 100 Mpc for face-on and edge-on events, respectively. However, since the field of view of an NIR telescope is not as large as that of the optical one~\citep{2015A&A...575A..25S}, the improvement in the source localization by the gravitational-wave observation is crucial.

\subsection{Comparison with different BNS models}

\begin{figure}
\begin{center}
 	 \includegraphics[width=0.95\linewidth]{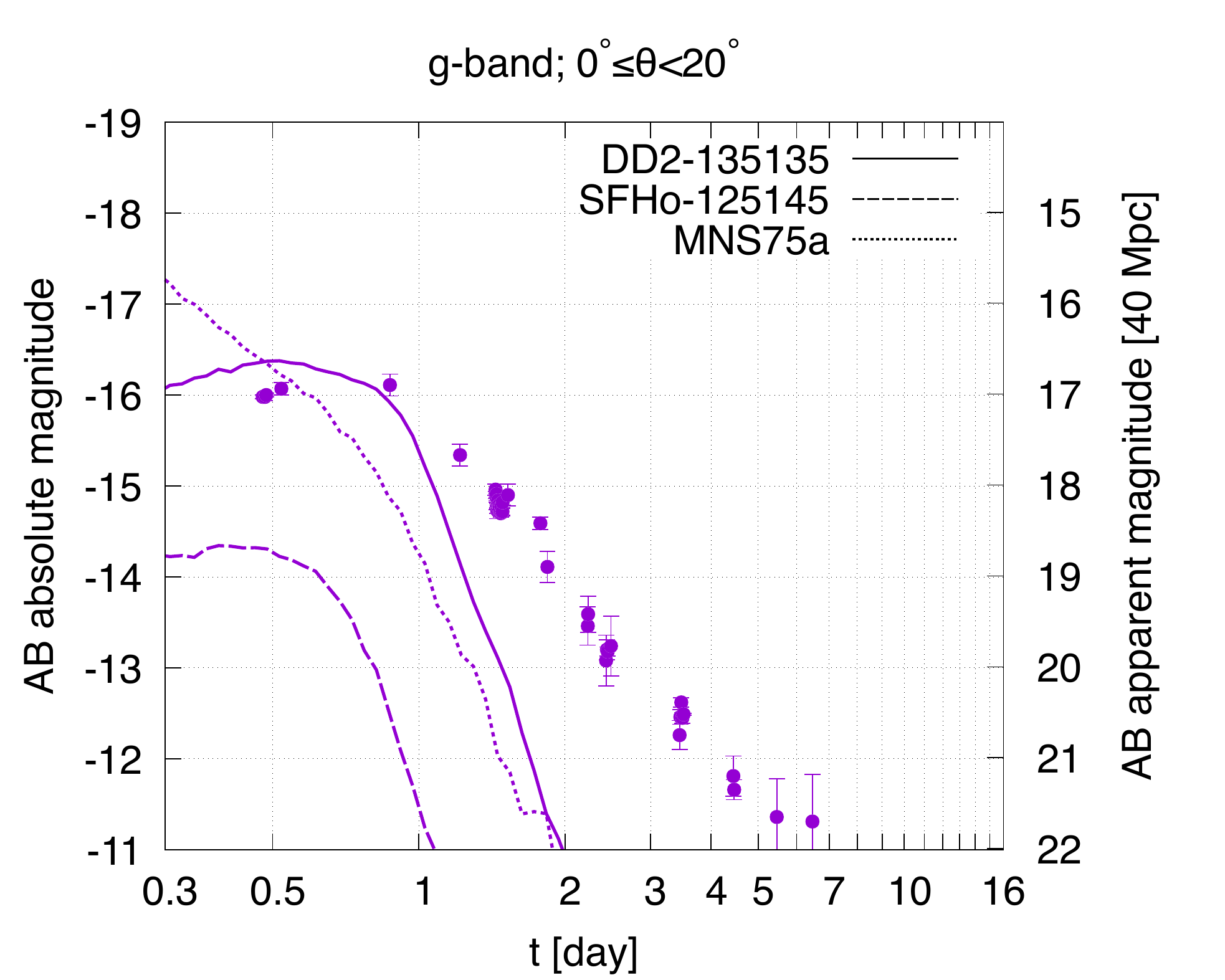}
 	 \includegraphics[width=0.95\linewidth]{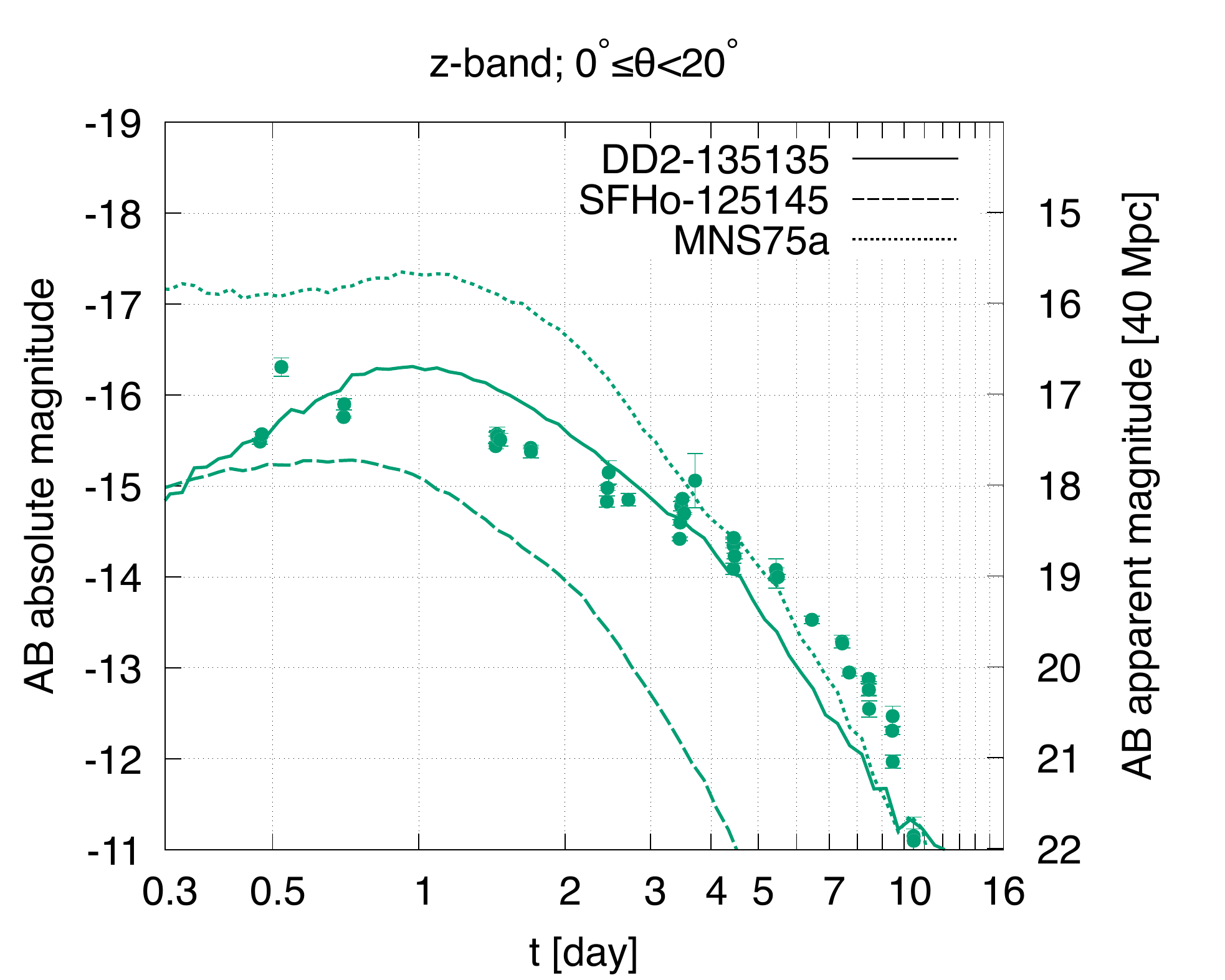}
 	 \includegraphics[width=0.95\linewidth]{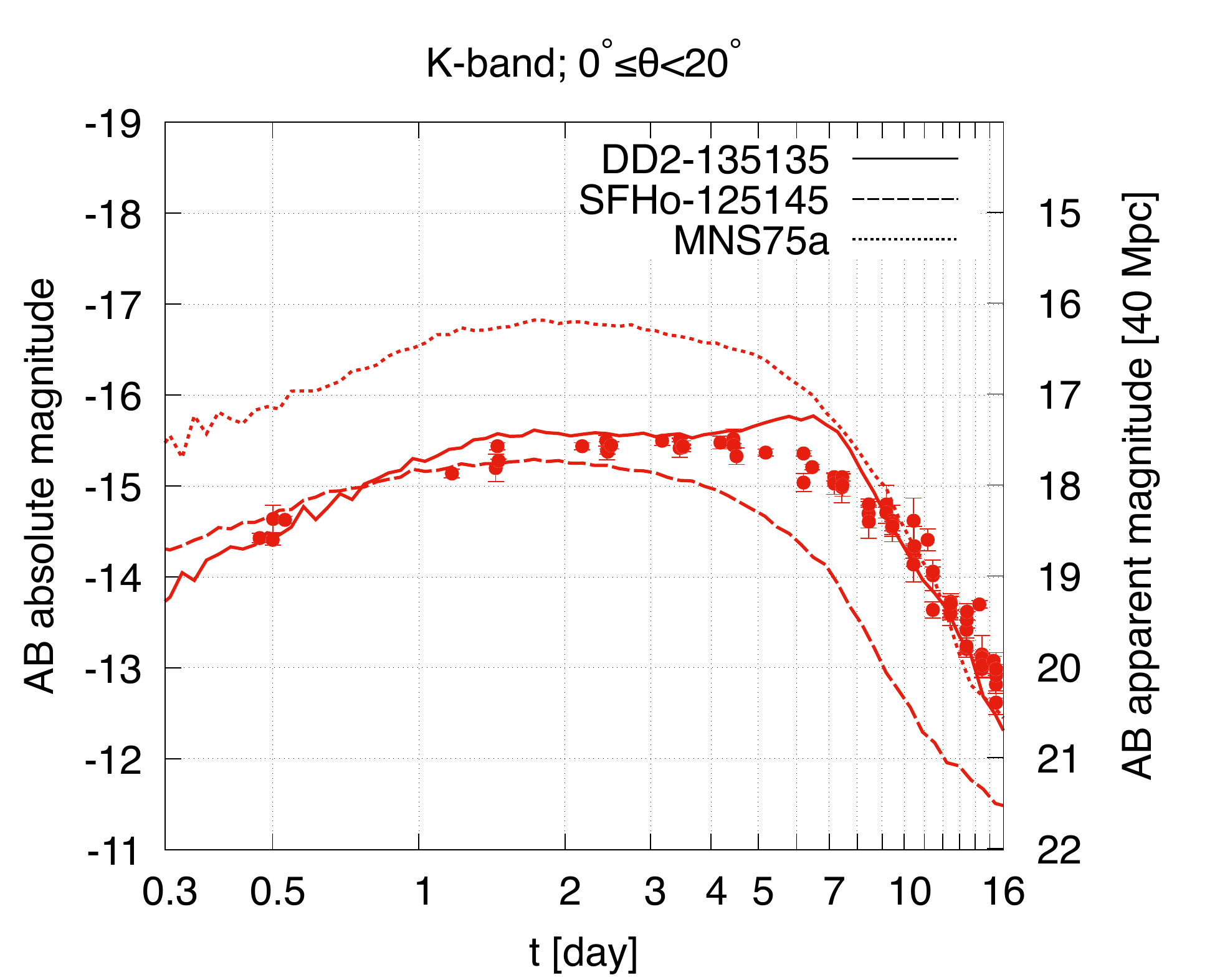}
 	 \caption{Comparison of the {\it gzK}-band light curves among the models in which remnant MNSs survive for a short time scale (the dashed curves; SFHo-125145) and for a long time scale (the solid curves; DD2-135135, ~\citet{Fujibayashi:2020dvr,Kawaguchi:2022bub}), and for the case in which significant magnetic dynamo effects are present in a long-surviving remnant MNS (the dotted curves; MNS75a, ~\citet{Shibata:2021xmo,Kawaguchi:2022bub}). The {\it g}, {\it z}, and {\it K}-band light curves are shown in the top, middle, and bottom panels, respectively.}
	 \label{fig:mag_comp}
  \end{center}
\end{figure}

Fig.~\ref{fig:mag_comp} compares the {\it gzK}-band kilonova light curves for the BNS models for which the remnant MNS survives for a short time scale (SFHo-125145) and for a long time scale (DD2-135135, ~\citet{Fujibayashi:2020dvr,Kawaguchi:2022bub}), and for the case that significant magnetic dynamo effects are hypothetically present in a long-surviving remnant MNS (MNS75a, ~\citealt{Shibata:2021xmo,Kawaguchi:2022bub}). The time scale for the emission to rapidly decline is much shorter for the model with a short-lived remnant MNS than those for the models associated with the formation of a long-lived MNS simply because the ejecta mass for the short-lived MSN models is smaller by a factor of 5--10 than that for the latter cases. The brightness at the peak is also high for the case with a long-lived MNS, and the difference is more significant in a shorter wavelength. 

As already mentioned, none of the merger models that result in a short-lived remnant MNS can explain the peak kilonova brightness of AT2017gfo observed in the {\it gz}-band, nor the brightness in the {\it K}-band in the late phase ($\gtrsim 5\,{\rm d}$). This is likely to be the case even if we consider a possible enhancement in the optical-band emission due to the modification in the ionization states by the non-LTE effects (see Appendix~\ref{app:nonLTE}). On the other hand, the kilonova model of a BNS that results in a long-surviving MNS (DD2-135135) reproduces the peak brightness in the optical wavelengths as well as the brightness and declining time scale in the NIR wavelengths, although a deviation from the observation is present in the optical wavelengths in the late phase ($t\gtrsim 2$ d)\footnote{Taking the non-LTE effects on the ionization populations into account may solve the tension; see~\citet{Kawaguchi:2020vbf,Kawaguchi:2022bub} for the discussion.}. This suggests that the formation of a short-lived remnant MNS is unlikely the case for GW170817 and the formation of an MNS which survives for a longer time scale ($\gtrsim 0.1\,{\rm s}$) is more likely from the viewpoint of kilonova light curves. However, for the case that the significant magnetic dynamo effects are present in the long-surviving remnant MNS (MNS75a), the kilonova emission will be significantly brighter than the observed data (see the light curves of MNS75a in Fig.~\ref{fig:mag_comp}). This suggests that the remnant MNS of GW170817 should have not survived for too long time (i.e., over the time scale of the dynamo magnetic-field amplification) if the magnetic dynamo effect played a significant role in the post-merger phase (see also the discussion below for the viewpoint of the nucleosynthesis yields).


\subsection{Approximate scaling law of kilonova light curves}

\begin{figure}
 	 \includegraphics[width=\linewidth]{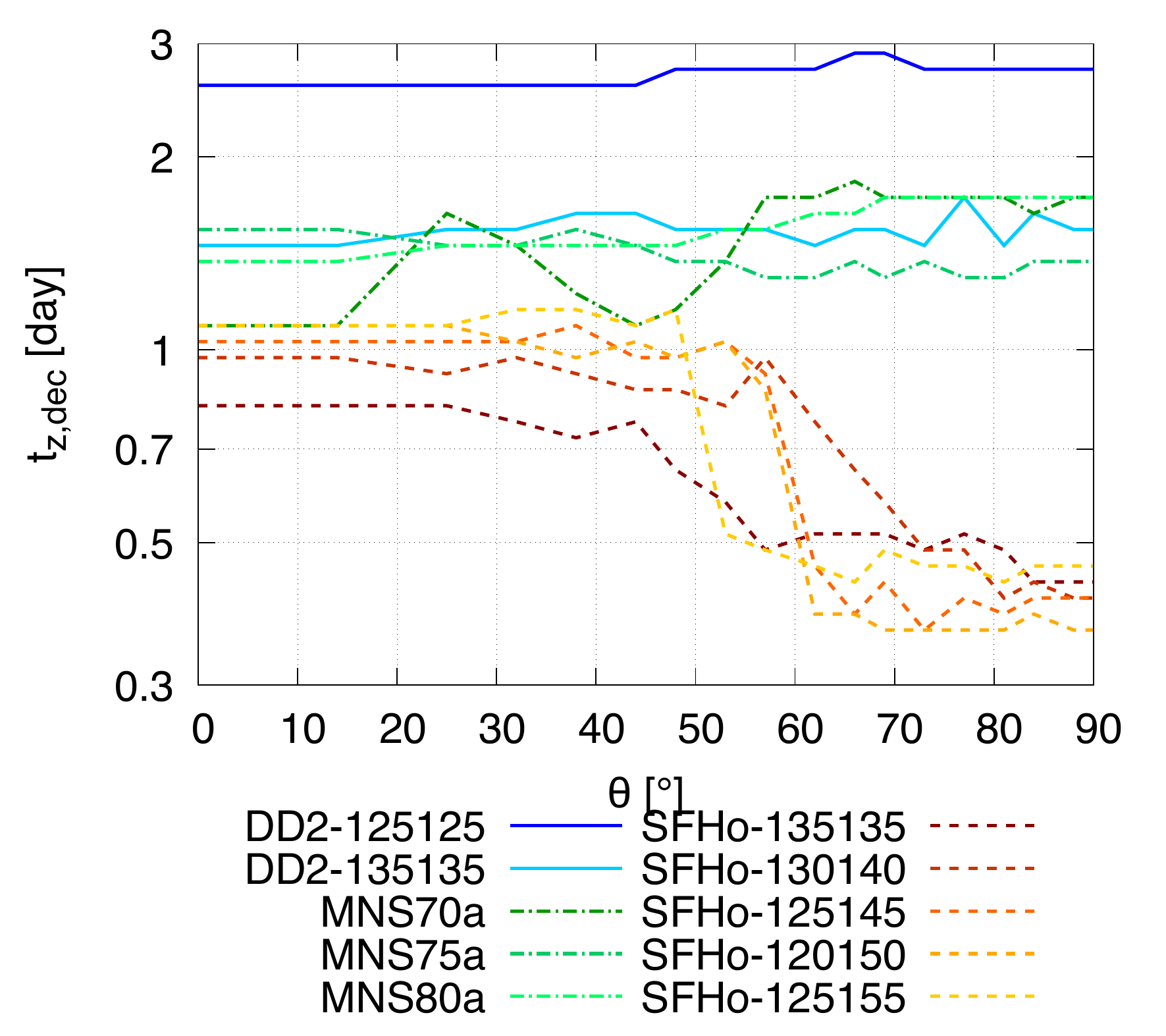}
 	 \includegraphics[width=\linewidth]{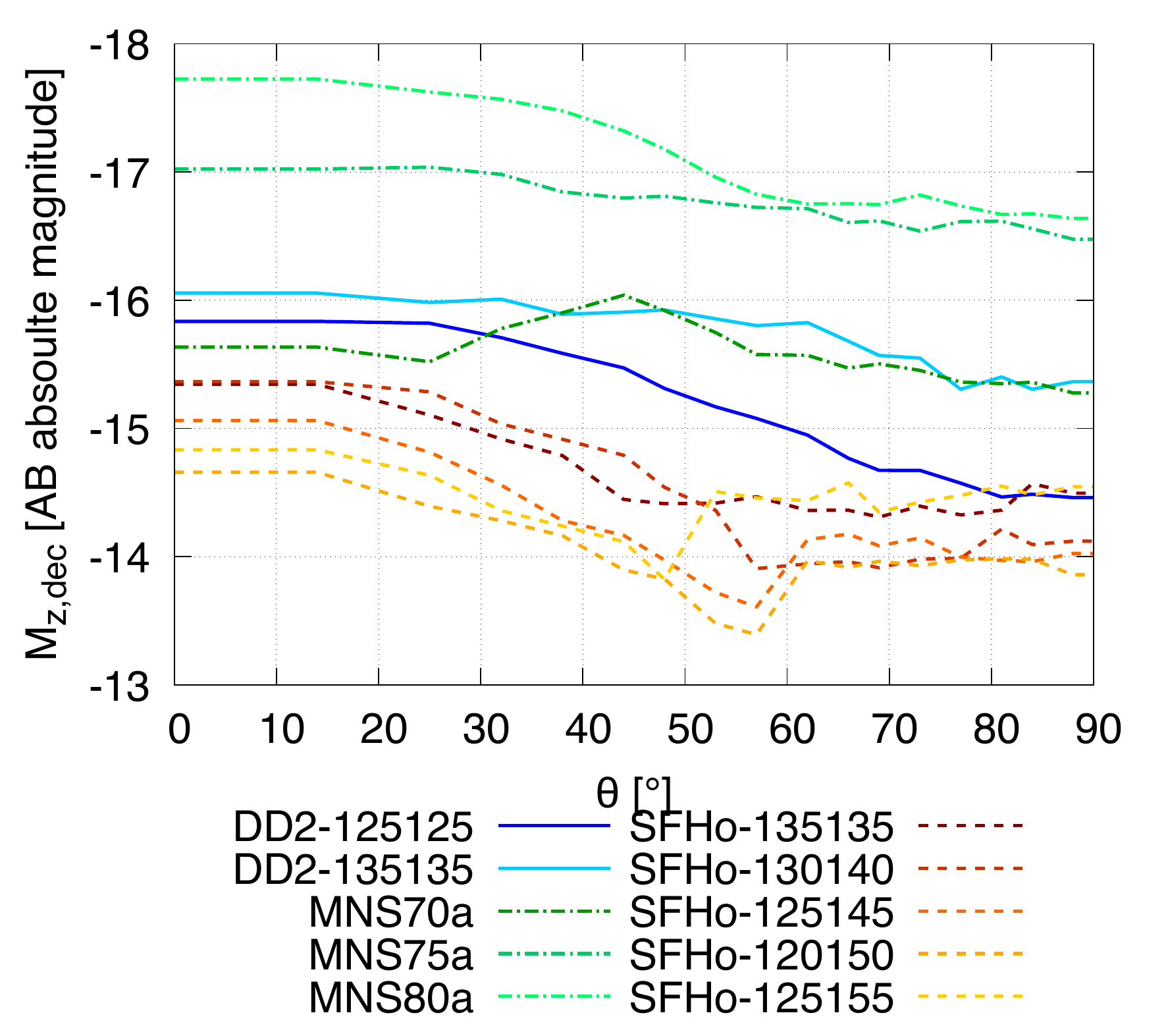}
 	 \caption{The time (top) and AB absolute magnitude (bottom) at which the decline power of the {\it z}-band magnitude, $dM_{\it z}/d{\rm log}_{10}t$, reaches to $2.5$ as functions of the viewing angle. The solid, dashed, and dash-dotted curves denote the cases in which the remnant MNS survives for a long period ($t\gg1\,{\rm s}$; DD2-125 and DD2-135 in~\citealt{Fujibayashi:2020dvr,Kawaguchi:2020vbf,Kawaguchi:2022bub}), the remnant MNS collapses to a BH in a short time ($t\lesssim 20\,{\rm ms}$; see Table~\ref{tb:model}), and the magnetic dynamo effects in the long-lived MNS are considered (MNS70a, MNS75a, and MNS80 in~\citealt{Shibata:2021xmo,Kawaguchi:2022bub}).}
	 \label{fig:decsl}
\end{figure}

\begin{figure}
 	 \includegraphics[width=\linewidth]{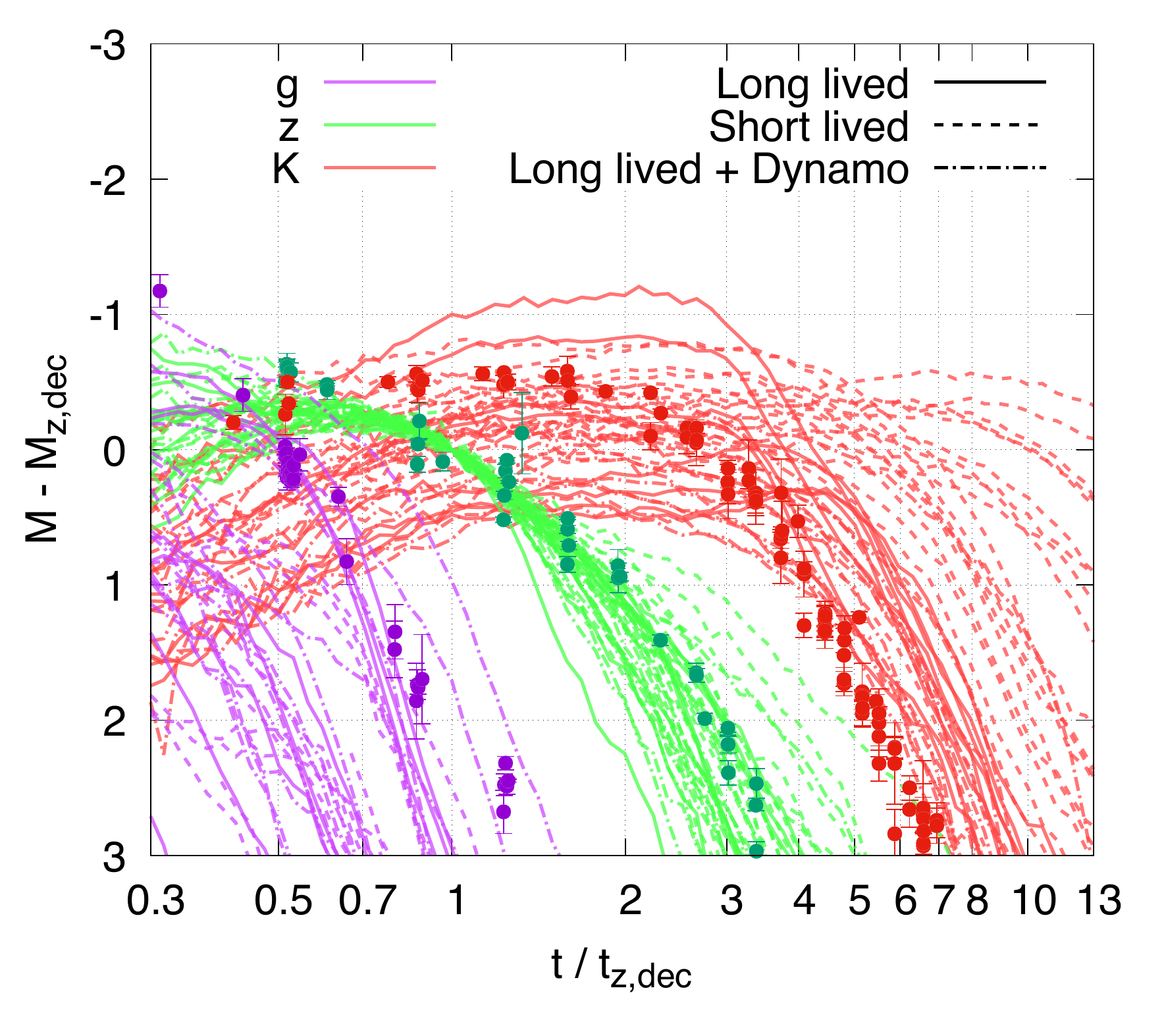}
 	 \caption{The {\it gzK}-band light curves for various models and viewing angles for which the time and magnitude are scaled by those at which the decline power of the {\it z}-band magnitude, $dM_{\it z}/d{\rm log}_{10}t$, reaches $2.5$. The solid, dashed, and dash-dotted curves denote the long-lived, short-lived, and long-lived dynamo cases, respectively, as in Fig.~\ref{fig:decsl}. The light curves observed from $0^\circ\le\theta\le20^\circ$, $28^\circ\le\theta\le35^\circ$, $59^\circ\le\theta\le64^\circ$, and $86^\circ\le\theta\le90^\circ$ are considered. The scaled observational data of AT2017gfo in the {\it gzK}-band taken from~\citet{Villar:2017wcc} are also plotted by circles with error bars.}
	 \label{fig:sl}
\end{figure}

While the peak brightness and the time scale of the emission differ among different BNS models and setups, Fig.~\ref{fig:mag_comp} implies that the shapes of the light curves as well as their relative brightness among different wavelengths share similar behaviour among the models. To examine this idea, we compare the {\it gzK}-band light curves for various models and viewing angles with the time and magnitude of each light curve being scaled by those at a certain reference time. 

For this purpose, we chose the reference time for each light curve to be the decline time of the {\it z}-band emission, $t_{\rm z,dec}$, defined as the time at which the decline power of the {\it z}-band magnitude, $dM_{\it z}/d{\rm log}_{10}t$, reaches $2.5$. Fig.~\ref{fig:decsl} shows the reference time and {\it z}-band magnitude as functions of the viewing angle for various kilonova models~\citep{Kawaguchi:2020vbf,Kawaguchi:2022bub}. The reference time and magnitude largely vary among the models and viewing angles. As expected from Fig.~\ref{fig:mag_comp}, the reference time and magnitude tend to be earlier and fainter, respectively, for the short-lived cases than the long-lived cases. The viewing-angle dependence is more pronounced for the short-lived cases, which reflects the fact that the dynamical component has a larger fraction in the total ejecta compared to the long-lived cases.

Fig.~\ref{fig:sl} compares the {\it gzK}-band light curves for various models and viewing angles, which are scaled with the reference time and $z$-band magnitude for each case. The {\it g}-band light curves show a large diversity among the models even after the scaling, for which we find no clear trend among the models and viewing-angles. On the other hand, although the reference time and magnitude largely vary among the models and viewing angles, the {\it K}-band light curves show relatively a less diversity after the scaling. In particular, the value of the {\it K}-band magnitude is always within $\approx 1$ mag relative to the value of the reference {\it z}-band magnitude for $0.6\lesssim t/t_{\rm z,dec}\lesssim 4$. We find that this is also the case for the {\it H} band. Hence, this suggests that the {\it HK}-band follow-up observation should be at least $1$ mag deeper than the value of the {\it z}-band reference magnitude and earlier than 4 times the reference time. 

Once the kilonova candidate is found and the decline time is determined by the {\it z}-band observation in a few days after the event, this approximate scaling law can be used as a guideline for the NIR follow-up observation by letting us know how rapid and how deep the observation should be. For example, let us suppose the case for which an EM candidate is found in the {\it z} band and $dM_{\it z}/d{\rm log}_{10}t$ reaches $2.5$ with the {\it z}-band magnitude being 20 mag at 1.5 d after the gravitational-wave trigger. Then, our approximate scaling-law suggests that the follow-up observation deeper than 21 mag within 6 d is at least needed not to miss the peak brightness of the {\it HK}-band counterparts.

Notably, the {\it K}-band emission tends to decline within $t/t_{\rm z,dec} \approx 5$--$10$ for the cases with a long-lived remnant MNS, while the {\it K}-band magnitude for the cases with a short-lived remnant MNS tends to keep the value close to the peak until a larger value of $t/t_{\rm z,dec}$. The observational data of AT2017gfo in the {\it gzK}-band scaled in the same way tend to follow the trend of the cases with a long-lived remnant MNS, which also supports our hypothesis that the remnant MNS for GW170817 did not collapse to a BH within a short time ($<20\,{\rm ms}$).

\section{Discussions}\label{sec:discuss}

We found that the kilonova light curves of a BNS of which the remnant MNS survives for a short time are too faint and last for a too short duration to explain the brightness of the optical and NIR observation of GW170817/AT2017gfo. This is primarily due to the smallness of ejecta mass. Instead, kilonova models of a BNS which results in a long-surviving MNS (DD2-135135) are more consistent with the observation. This indicates that the remnant MNS of GW170817 might not have collapsed within a short time ($\lesssim 20\,{\rm ms})$ but survived for a longer time ($\gtrsim 0.1\,{\rm s}$). On the other hand, our previous study~\citep{Kawaguchi:2022bub} indicated that, if the dynamo effects play a significant role for an efficient amplification of magnetic fields in a long-lived remnant MNS, the kilonova as well as the synchrotron emission stemming from the interaction between the ejecta fast tail and inter-stellar medium becomes too bright to be consistent with the EM observations associated with GW170817 (see also the discussion in~\citet{Sarin:2022wby}). Hence, the remnant MNS should have collapsed to a BH within the dynamo time scale of the magnetic-field growth, or the dynamo effect in the post-merger phase was subdominant.

\begin{figure}
\begin{center}
 	 \includegraphics[width=1.\linewidth]{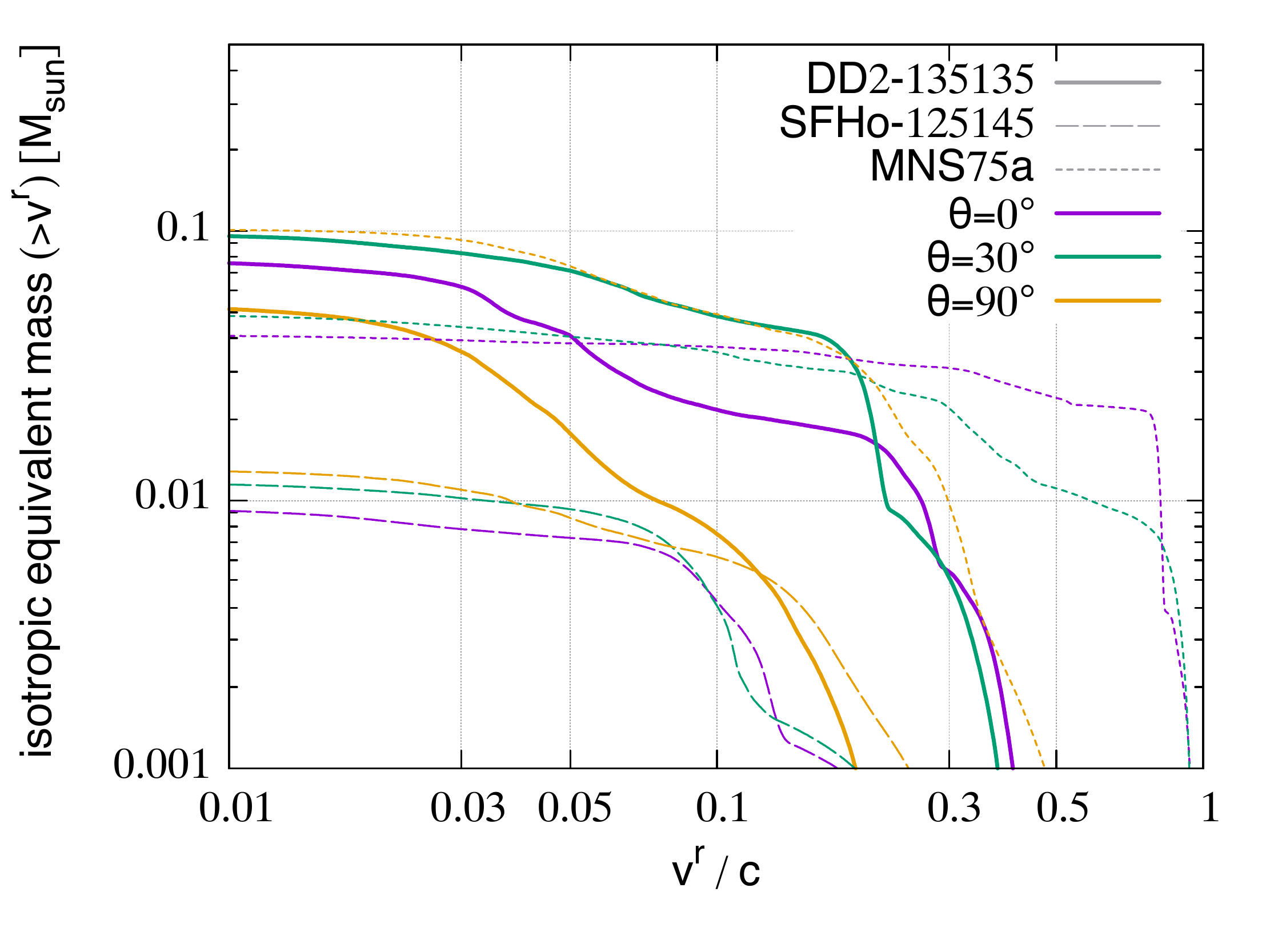}
 	 \caption{Isotropic equivalent ejecta mass for various models and latitudinal angles. The solid, dashed, and dash-dotted curves denote the long-lived, short-lived, and long-lived dynamo cases, respectively, as in Fig.~\ref{fig:decsl}. The purple, green, and orange curves denote the results for $\theta=0^\circ$,  $30^\circ$, and $90^\circ$, respectively.}
	 \label{fig:isomass}
  \end{center}
\end{figure}

We find that the mass distribution of the ejecta in the polar region for the long-lived case is also compatible with the required property of the fast blue component, for which the origin is often discussed to be mysterious~\cite[e.g.,][]{Kasliwal:2017ngb,Cowperthwaite:2017dyu,Kasen:2017sxr,Villar:2017wcc,Waxman:2017sqv,Kawaguchi:2018ptg,Kawaguchi:2019nju,Bulla:2019muo,Almualla:2021znj,Kedia:2022onl,Bulla:2022mwo}. Fig.~\ref{fig:isomass} shows the isotropic equivalent ejecta mass, $M^{\rm iso}_{\rm eje}(v^r,\theta)$, for various models and latitudinal angles, which is defined by
\begin{align}
    M^{\rm iso}_{\rm eje}(v^r,\theta)=4\pi\int_{>v^r} \rho(r,\theta) r^2 dr,
\end{align}
where $\rho$ denotes the rest-mass density. For the case of long-lived MNS formation (DD2-135135), the polar value of $M^{\rm iso}_{\rm eje}$ for $v^r\gtrsim0.2\,c$ is larger than $10^{-2}\,M_\odot$. This matches the property of the ejecta which is required to explain the luminosity and photo-spheric velocity of the blue component in AT2017gfo (see also~\citealt{Just:2023wtj} for similar findings). Such a polar ejecta component is originated from the dynamical ejecta component and the post-merger ejecta component of which the velocity is enhanced by neutrino-radiation from the MNS. While the spectral analysis with the non-LTE effects being taken into account is needed for a more quantitative argument, our finding suggests that the photo-spheric velocity of the blue component can be naturally explained by the setup obtained by NR simulations.

Fig.~\ref{fig:isomass} suggests that the presence of the diversity in the evolution of photos-spheric velocities reflects the different types of the MNS evolution. For the case of short-lived MNS formation (SFHo-125145), the value of $M^{\rm iso}_{\rm eje}$ only reaches $10^{-2}\,M_\odot$ for $v_r<0.05\,c$, simply reflecting the smallness of the ejecta mass. This suggests that the photo-spheric velocity of the short-lived case is $\lesssim 0.05\,c$ for $t\gtrsim 1$ d. On the other hand, the result of MNS75a shows that an applicable amount of ejecta is distributed in the very high velocity components. This is due to the acceleration of ejecta in the presence of significant magnetic dynamo effects in the long-lived MNS, and the photo-spheric velocity of $>0.8\,c$ is expected be observed in the early phase of emission for such a case.

\begin{figure}
\begin{center}
 	 \includegraphics[width=1.\linewidth]{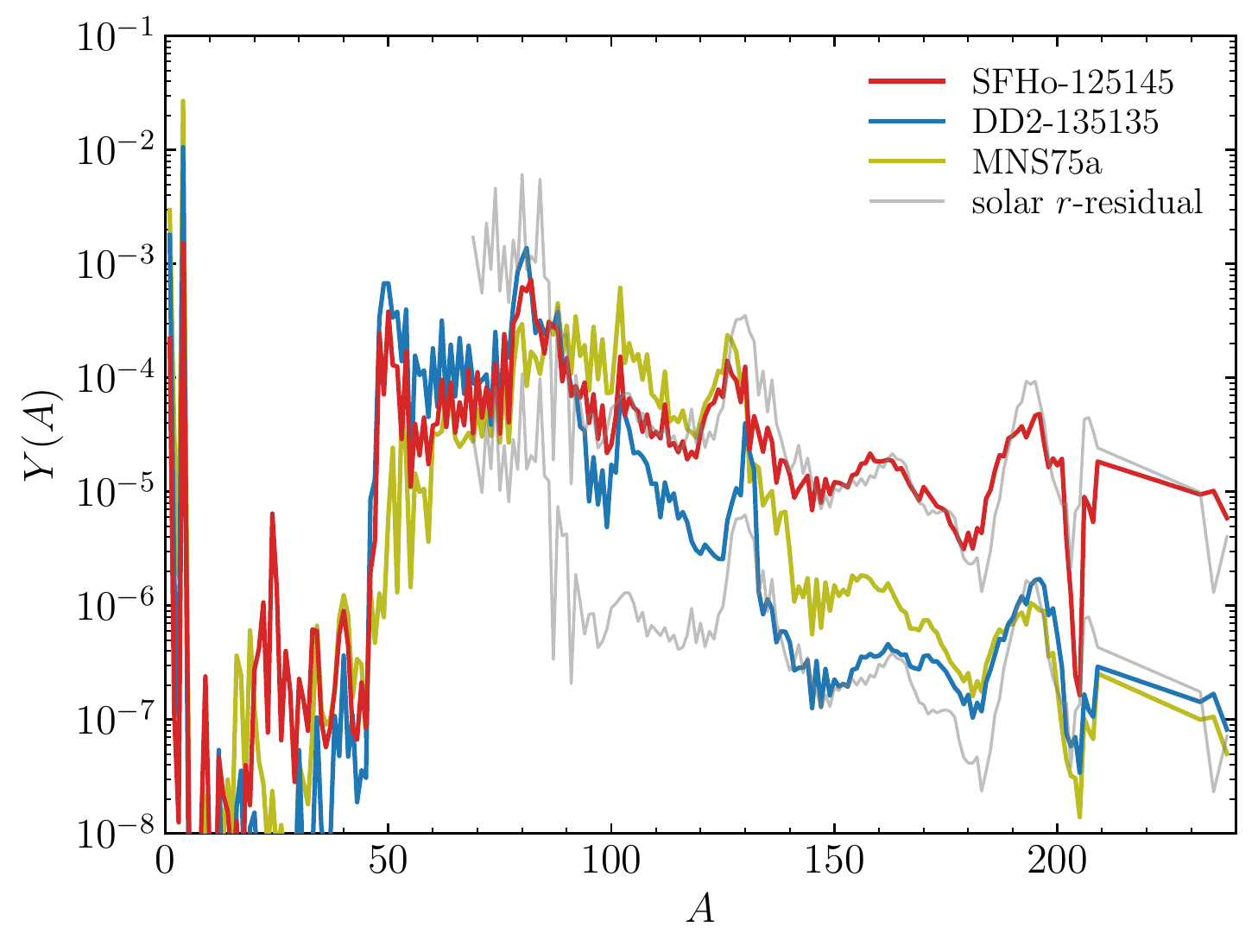}
 	 \caption{Comparison of nucleosynthesis yields among the cases in which remnant MNSs survive for a short time (red; SFHo-125145) and for a long time (blue; DD2-135135, ~\citealt{Fujibayashi:2020dvr}), and for the case in which significant magnetic dynamo effects are present in a long-surviving remnant MNS (olive; MNS75a, ~\citealt{Shibata:2021xmo}). The $r$-process residuals to the solar system abundances \citep{Lodders2009} are also shown by gray curves, which are scaled to match the abundance of $^{153}$Eu for SFHo-125145 as well as that for DD2-135135.}
	 \label{fig:abun_comp}
  \end{center}
\end{figure}

As described above, the BNS that results in a long-lived MNS is more likely the case for GW170817 than the BNS that results in a short-lived remnant MNS from the viewpoint of kilonova light curves. However, the calculated nucleosynthesis yields for such long-lived MNS cases (DD2-135135 and MNS75a in Fig.~\ref{fig:abun_comp}) exhibit overproduction of the nuclei between the first and second $r$-process abundance peaks ($A \sim 80$--130) when compared to the solar $r$-process abundances (see also~\citealt{Fujibayashi:2020dvr,Shibata:2021xmo} for the details, and~\citealt{Just:2023wtj} for similar results). This fact suggests that such long-lived MNSs should not be the major outcomes of BNSs that merge in a Hubble time if the dominant sources of $r$-process elements are BNS mergers. This implies that GW170817 may not be a typical type of BNS mergers in the universe. 

However, we should note that the total nucleosynthesis yields can be sensitive to the setups and physical ingredients of the numerical simulation. A latest work suggests that a more self-consistent magnetohydrodynamics treatment of angular momentum transfer could result in more production of elements heavier than the first $r$-process peak in the post-merger ejecta~\citep{Kiuchi:2022nin}. Hence, there may still exist a room that both the observation of GW170817 and the robustness of the solar abundance pattern~\citep{Cowan:2019pkx} can be explained by some configuration of a BNS, while we should remind that the presence of an MNS which survives for a long time scale ($t>1\,{\rm s}$) with significant dynamo effects is unlikely the case of GW170817 as discussed above. For example, a BNS which results in a remnant MNS with significant dynamo effects but collapses to a BH at $O(0.1)$\,s can be a plausible model for interpreting GW170817 from this point of view. 

For the BNS resulting in a short-lived MNS, the kilonova emission lasts over a time scale appreciably shorter than that of GW170817/AT2017gfo, in particular for the optical band. This implies that for detecting kilonovae of this type, we need observation earlier than that for AT2017gfo. This is in particular the case for a large value of $\theta$. It is also likely that the optical light curves could be more easily hidden by the afterglow light curves of GRBs for the small value of $\theta$. Hence, the NIR light curves may be the primary target of the observation in the simultaneous detection of a GRB.

\begin{figure}
\begin{center}
 	 \includegraphics[width=1.\linewidth]{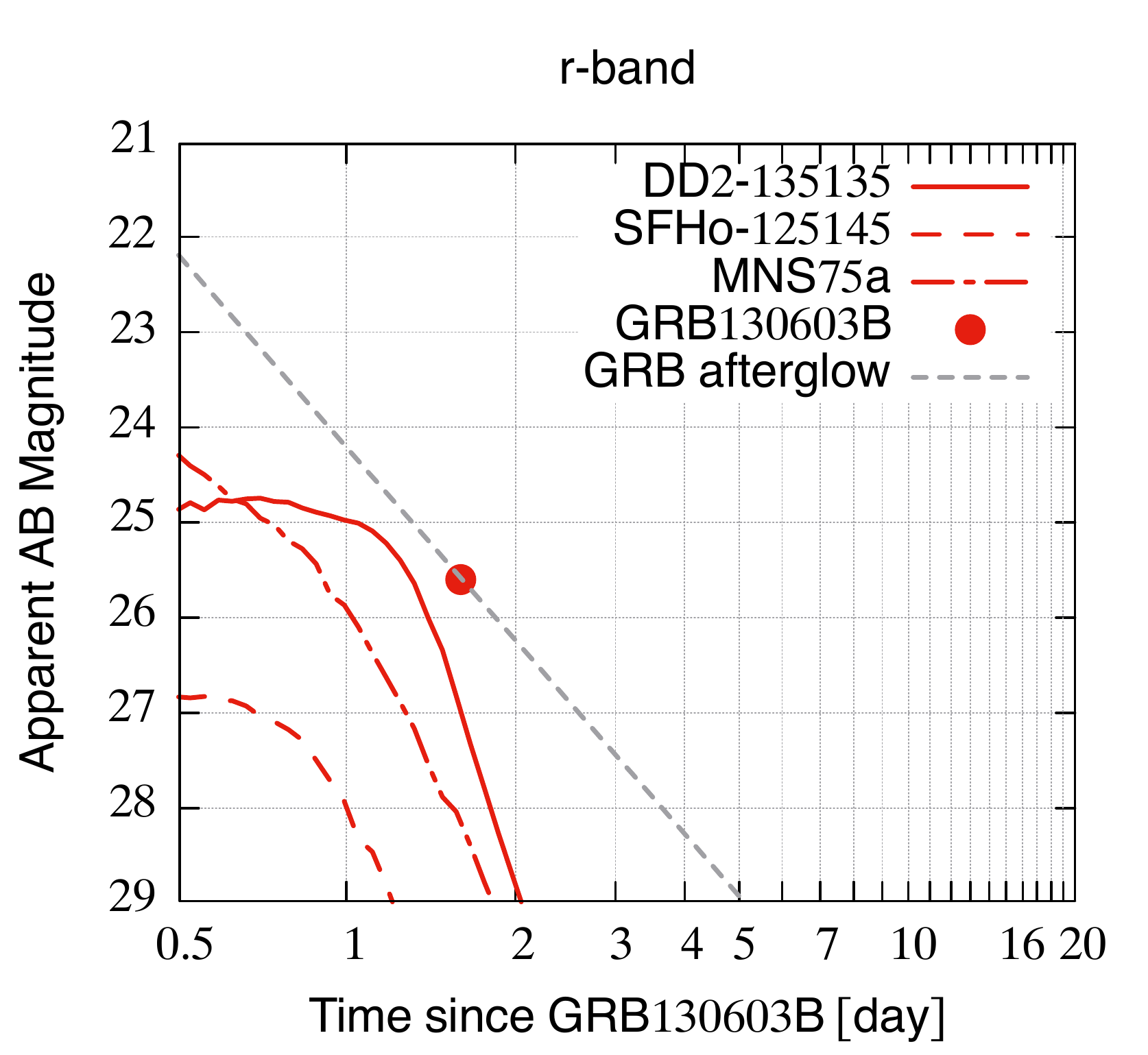} \\  
 	 \includegraphics[width=1.\linewidth]{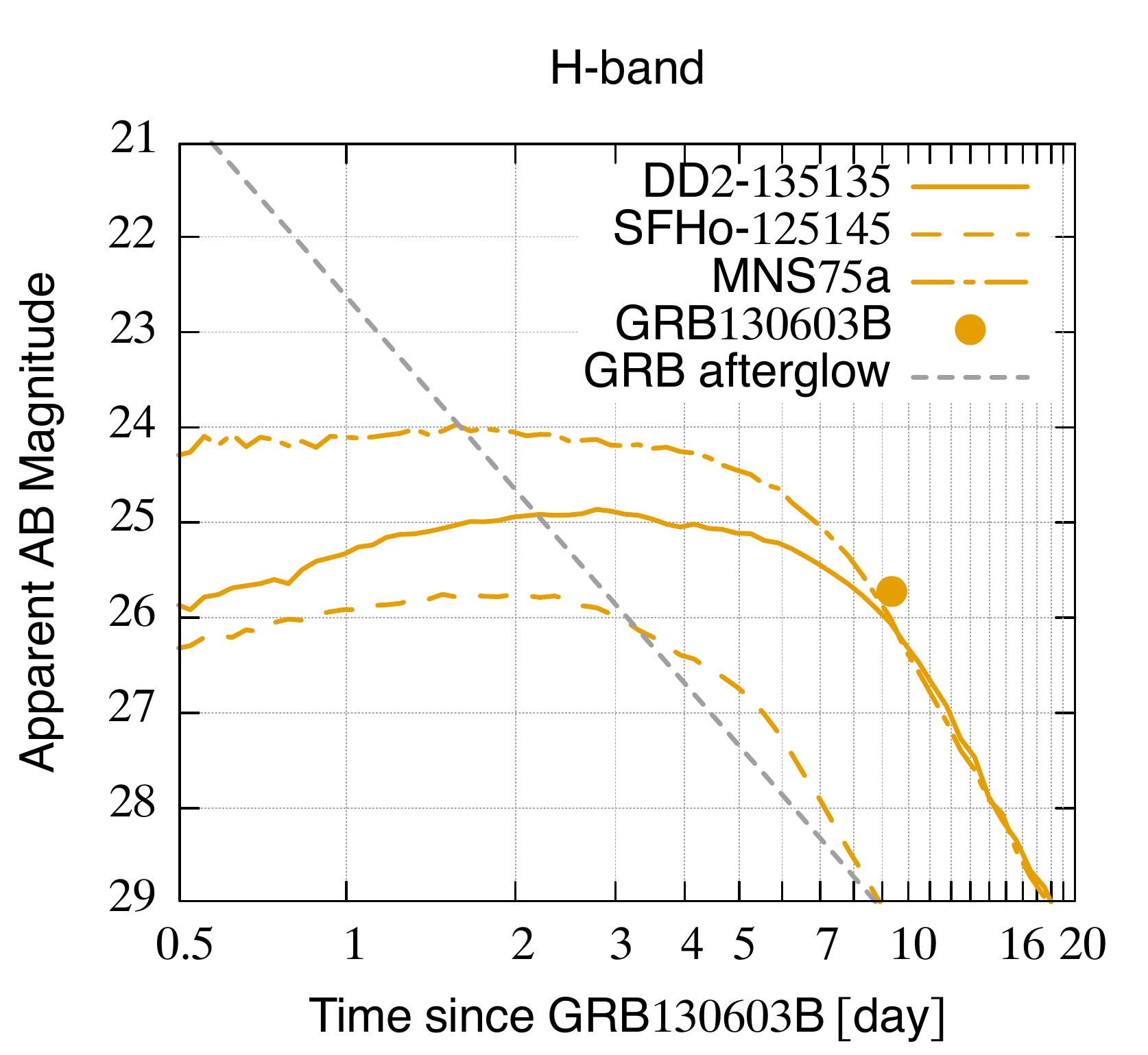}
 	 \caption{Comparison between the optical and NIR observation in GRB130603B and various kilonova models. The {\it r}- and {\it H}-band light curves in the observer frame are calculated by employing the redshift value of the source~\citep[$z=0.356$, ][]{2013GCN.14744....1T,Cucchiara:2013vda}. The solid, dashed, and dash-dotted curves denote the long-lived, short-lived, and long-lived dynamo cases, respectively, as in Fig.~\ref{fig:decsl}. The gray dashed lines denote the GRB afterglow light curves. The observational data points for the $r$- and $H$-band magnitudes (circles) in the GRB130603B observation and the GRB afterglow model light curves are taken from~\citet{Tanvir:2013pia}.}
	 \label{fig:grb130603b}
  \end{center}
\end{figure}

In fact, the comparison of our model light curves with the observation of  GRB130603B~\citep{Berger:2013wna,Tanvir:2013pia}, with which a plausible kilonova candidate is associated, indicates that the {\it r}-band emission for the case of short-lived MNS formation (SFHo-125145) is likely hidden by the afterglow emission (see Fig.~\ref{fig:grb130603b}). The brightness in the {\it H} band for the case of short-lived MNS formation is also at most only comparable to that of the afterglow emission. Hence, the progenitor of GRB130603B was unlikely to be a BNS which results in the formation of a short-lived remnant MNS assuming the excess in the {\it H} band is due to the kilonova emission. This also indicates that GRB-associated kilonovae from BNSs leading to short-lived MNS formation could be missed by being entirely hidden by the afterglows, which should result in a number of simultaneous detection of gravitational waves with short GRBs but lack of kilonovae in future. Indeed a statistical study shows that there are a substantial fraction of previous short GRBs that are not associated with kilonovae~\citep{Troja:2023cev}.

As the brightness of AT2017gfo is known to be broadly comparable with the optical and NIR counterparts of GRB130603B~\citep{Rossi:2019fnm}, the kilonova model light curves for the cases of long-lived MNS formation (DD2-135135 and MNS75a) are also consistent with the observation of GRB130603B; while the {\it r}-band emission is hidden by the afterglow emission, the {\it H} band emission for the long-lived cases is brighter than the afterglow emission, and is consistent with the observed excess. This suggests that the progenitor of GRB130603B is likely to be a BNS which results in the formation of a MNS that survives more than $\sim 10\,{\rm ms}$.

\begin{figure*}
\begin{center}
   \hspace{-0.5cm}
 	 \includegraphics[width=.55\linewidth]{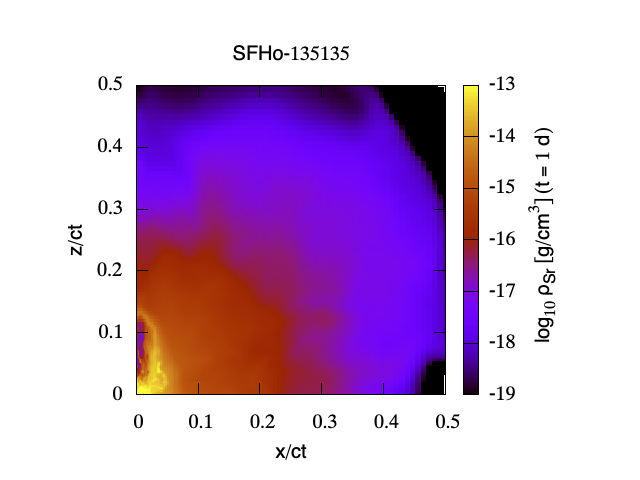} 
   \hspace{-2.cm}
 	 \includegraphics[width=.55\linewidth]{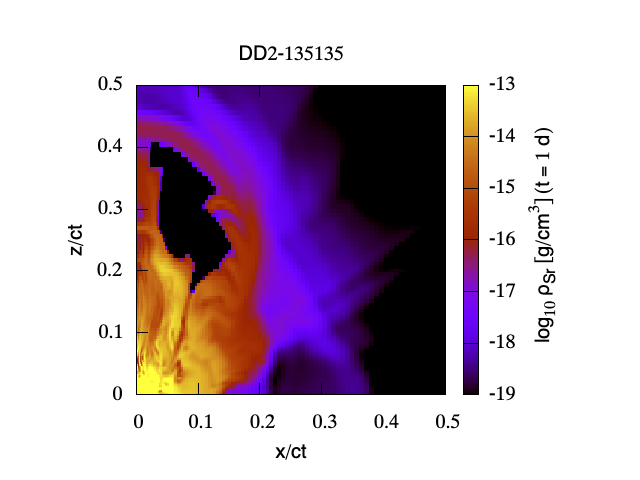}
 	 \caption{Sr mass density profiles for SFHo-135135 and DD2-135135~\citep{Fujibayashi:2020dvr,Kawaguchi:2022bub} at $t=1\,{\rm d}$.}
	 \label{fig:snabun}
  \end{center}
\end{figure*}

In~\citet{2019Natur.574..497W,Domoto:2021xfq,Domoto:2022cqp,Gillanders:2022opm}, spectral features observed in the data of AT2017gfo are interpreted as the p-Cygni profiles by Sr (note, however, ~\citet{Perego:2020evn,Tarumi:2023apl} suggested that the spectral features could be also well interpreted by the absorption lines by He if non-LTE effects are considered). Recently,~\citet{Sneppen:2023vkk} performed a more detailed analysis for those spectral features, and show that the Sr distribution of the ejecta should have nearly spherical morphology. Fig.~\ref{fig:snabun} shows the Sr mass density profiles at $t=1\,{\rm d}$ for SFHo-135135 and DD2-135135~\citep{Fujibayashi:2020dvr,Kawaguchi:2022bub}. The Sr distribution with the velocity larger than $0.15\,c$ approximately exhibits a spherical morphology for SFHo-135135. On the other hand, the Sr distributions for DD2-135135 as well as the low-velocity part ($<0.15\,c$) for SFHo-135135 show mildly prolate shapes. These aspherical features, which are in broad agreement with the results of~\citet{Just:2023wtj}, are inconsistent with the implication of~\citet{Sneppen:2023vkk}. Detailed quantitative spectral analysis taking into account various uncertainties is nevertheless needed to clarify how severe the current tension from the observational implication is, which we leave it for a future task.

\section*{Acknowledgements}
KK thanks Masaomi Tanaka and Eli Waxman for the valuable discussions. We also thank Kenta Hotokezaka for helpful discussions. Numerical computation was performed on Yukawa21 at Yukawa Institute for Theoretical Physics, Kyoto University and the Sakura, Cobra, Raven clusters at Max Planck Computing and Data Facility. The simulations were performed on Fugaku provided by RIKEN through the HPCI System Research Project (Project ID: hp220174, hp230084), and the Cray XC50 at CfCA of the National Astronomical Observatory of Japan. ND acknowledges support from Graduate Program on Physics for the Universe (GP-PU) at Tohoku University. This work was supported by Grant-in-Aid for Scientific Research (JP20H00158, JP21K13912, JP23H04900, 22KJ0317, 23H01772) of JSPS/MEXT.



\bibliographystyle{mnras}



\appendix

\section{Estimate of uncertainties due to non-LTE Effects}\label{app:nonLTE}

\begin{figure*}
\begin{center}
 	 \includegraphics[width=.45\linewidth]{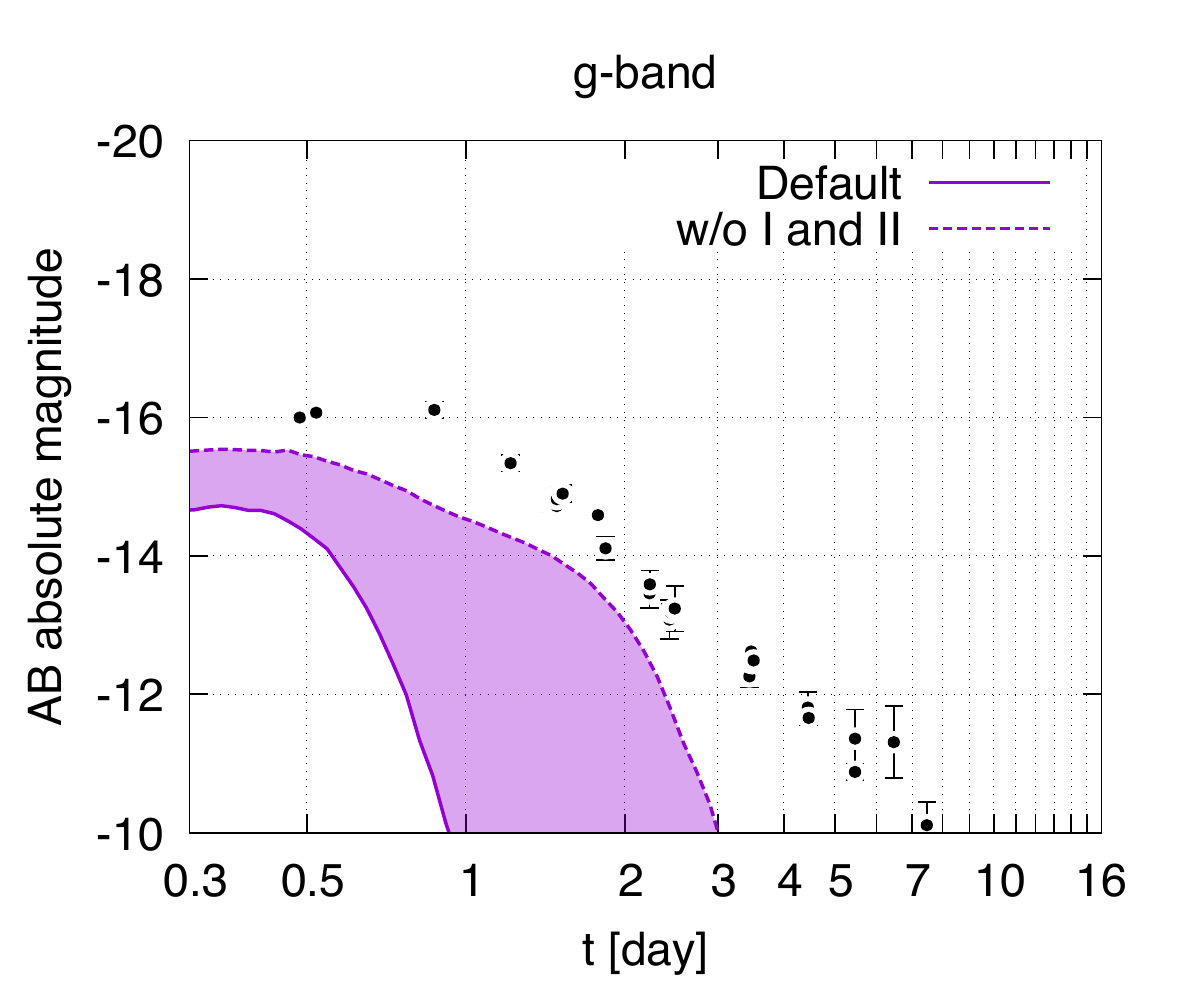}   
 	 \includegraphics[width=.45\linewidth]{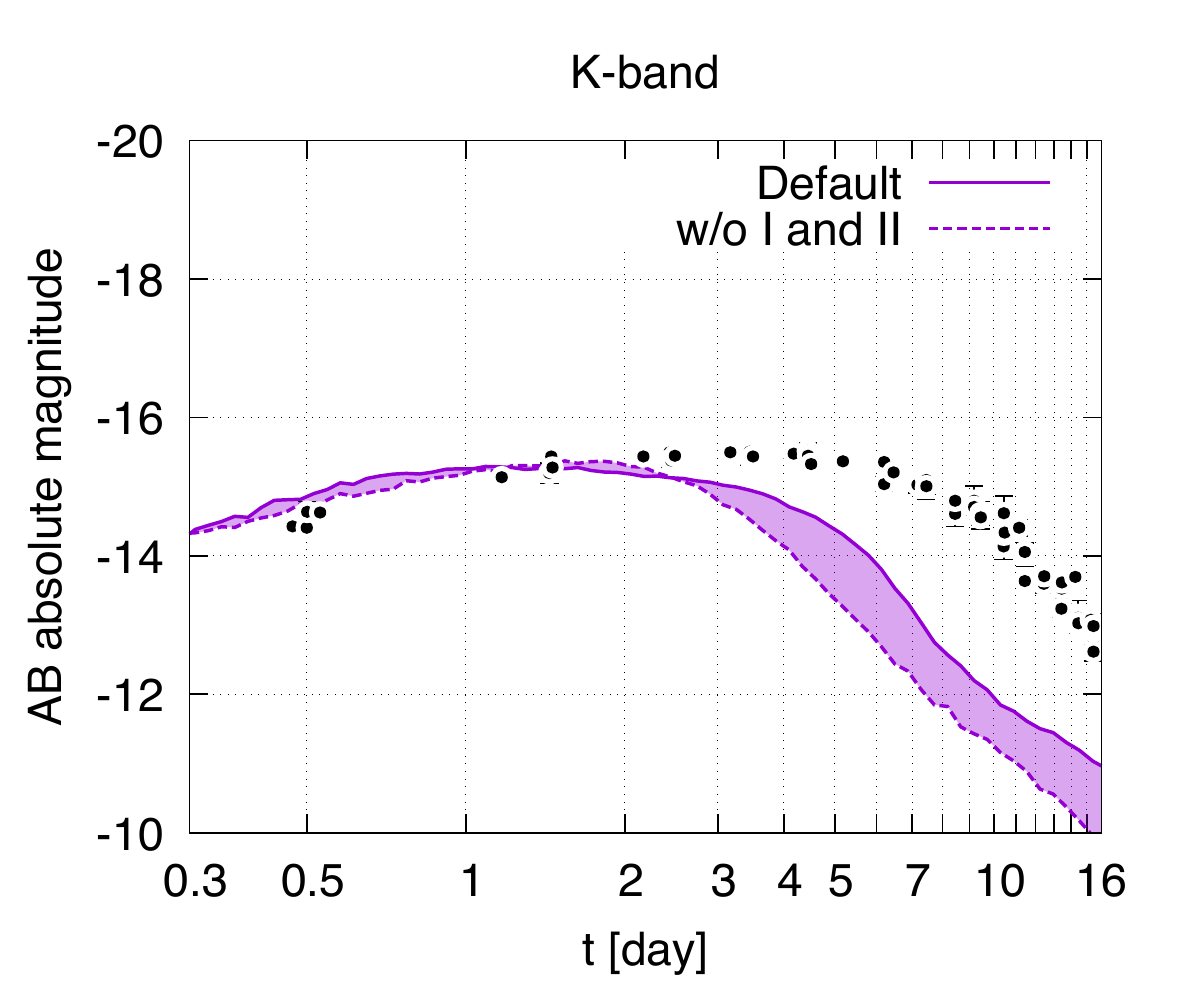}\\
 	 \includegraphics[width=.45\linewidth]{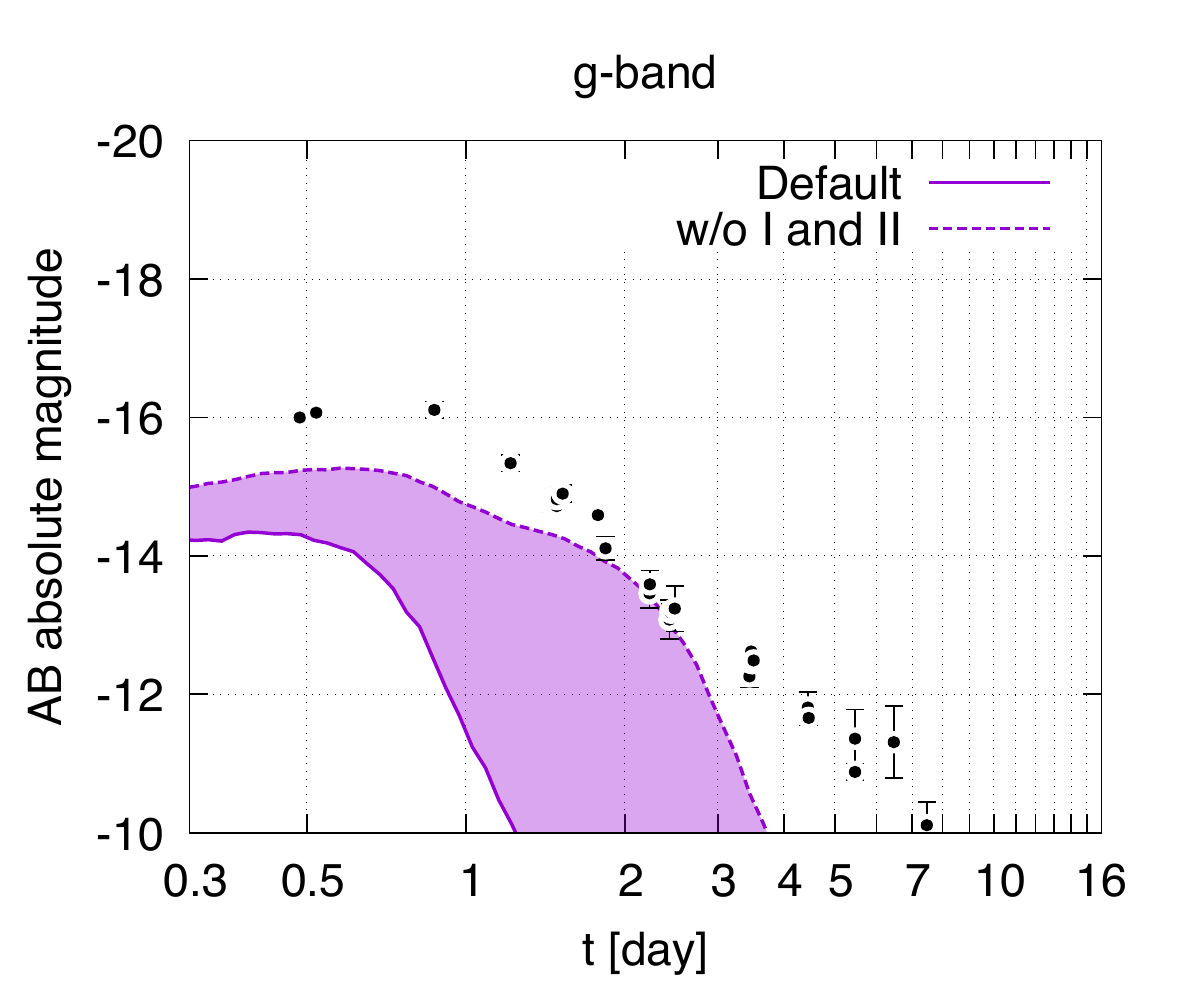}   
 	 \includegraphics[width=.45\linewidth]{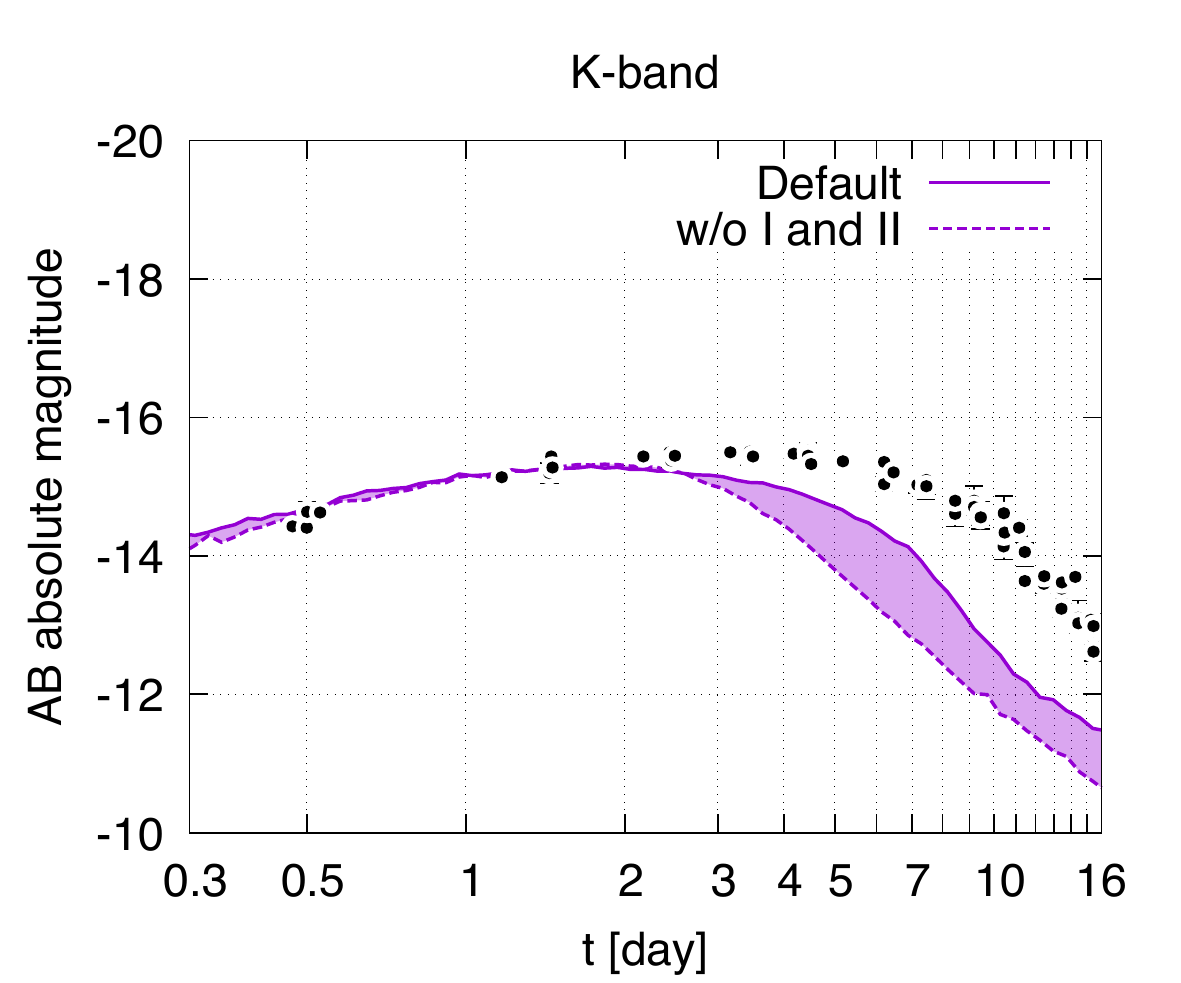}
 	 \caption{Comparison of the {\it gK}-band light curves observed from $0^\circ\leq\theta\leq20^\circ$ for models in which the presence of both neutral and first ionized atoms (``w/o I and II''; dotted curves) are prohibited. The light curves shown in the solid curves (``Default'') are the results with the default setting and are the same as those in Fig.~\ref{fig:mag}. The left and right panels show the results of the {\it g} and {\it K}-band light curves, respectively. The top and bottom panels show the results for models SFHo-135135 and SFHo-125145, respectively. The data points denote the observational data of AT2017gfo taken from~\citet{Villar:2017wcc} for a hypothetical distance to the source of 40\,Mpc.}
	 \label{fig:nonlte}
  \end{center}
\end{figure*}

In our present radiative-transfer simulations, the LTE condition is assumed to determine the ionization/excitation populations of atoms. This assumption can be invalid for the later phase of kilonova emission, at which the ionization of the atoms caused by radioactive decay becomes more significant than the recombination of ions~\citep{Hotokezaka:2021ofe,Kawaguchi:2020vbf}. \citet{Hotokezaka:2021ofe,Pognan2022MNRAS} indicate that such non-LTE effects may suppress the neutral and first ionized ions in the outer part of the ejecta even in the earlier phases. Such modifications in the ionization/excitation populations of atoms can have a great impact to the opacity and resultant light curves.

Because computing the ionization/excitation population is challenging due to its computational complexity and lack of the atomic data for $r$-process elements (see~\citealt{Hotokezaka:2021ofe,Pognan2022MNRAS}), we here provide qualitative estimates for the impacts of the non-LTE effects to the kilonova light curves following the same prescription which we applied in our previous studies~\citep{Kawaguchi:2020vbf,Kawaguchi:2022bub}; we perform the radiative transfer simulations with a hypothetical setup in which both neutral and first ionized atoms are artificially forced to be ionized to the second ionization states. Note that this prescription is applied to whole ejecta, including high-density regions for simplicity. 

Fig.~\ref{fig:nonlte} shows the {\it g} and {\it K}-band light curves for models SFHo-135135 and SFHo-125145 obtained with these hypothetical setups. As is also found in~\citet{Kawaguchi:2020vbf,Kawaguchi:2022bub}, the emission in the optical wavelengths is enhanced by artificially increasing the ionization degrees. Yet, the brightness of the {\it g}-band emission is not high enough to explain the brightness of AT2017gfo. The emission in the NIR wavelengths in the late phase becomes even fainter and more inconsistent with the observation. These results indicate that the BNS that results in a short-lived remnant MNS is, at least in our studied range of binary configurations, likely to be different from the BNS of GW170817.
\label{lastpage}
\end{document}